\documentclass[conference]{IEEEtran}
\IEEEoverridecommandlockouts
\usepackage{cite}
\usepackage{amsmath,amssymb,amsfonts}
\usepackage{algorithmic}
\usepackage[linesnumbered,ruled]{algorithm2e}
\usepackage{graphicx}
\usepackage{textcomp}
\usepackage{xcolor}
\usepackage{multirow}
\usepackage{xcolor,colortbl}
\definecolor{green}{rgb}{0.5,0.9,0.0}
\newcommand{\optimalthreeglatency}{\cellcolor{green}14.3}  
\newcommand{\optimalthreegenergy}{\cellcolor{green}13.1}  
\newcommand{\optimalfourglatency}{\cellcolor{green}5.2}  
\newcommand{\optimalfourgenergy}{\cellcolor{green}9.8}  
\newcommand{\optimalwifilatency}{\cellcolor{green}2.4}  
\newcommand{\optimalwifienergy}{\cellcolor{green}4.8}  

\def\BibTeX{{\rm B\kern-.05em{\sc i\kern-.025em b}\kern-.08em
    T\kern-.1667em\lower.7ex\hbox{E}\kern-.125emX}}

\renewcommand\baselinestretch{0.8805}
\begin{document}

\title{Towards Collaborative Intelligence Friendly Architectures for Deep Learning\\
{\footnotesize}
\thanks{}
}

\author{\IEEEauthorblockN{Amir Erfan Eshratifar}
\IEEEauthorblockA{\textit{Department of Electrical Engineering} \\
\textit{University of Southern California}\\
Los Angeles, USA \\
eshratif@usc.edu}
\and
\IEEEauthorblockN{Amirhossein Esmaili}
\IEEEauthorblockA{\textit{Department of Electrical Engineering} \\
\textit{University of Southern California}\\
Los Angeles, USA \\
esmailid@usc.edu}
\and
\IEEEauthorblockN{Massoud Pedram}
\IEEEauthorblockA{\textit{Department of Electrical Engineering} \\
\textit{University of Southern California}\\
Los Angeles, USA \\
pedram@usc.edu}
}

\maketitle

\begin{abstract}
Modern mobile devices are equipped with high-performance hardware resources such as graphics processing units (GPUs), making the end-side intelligent services more feasible. Even recently, specialized silicons as neural engines are being used for mobile devices. However, most mobile devices are still not capable of performing real-time inference using very deep models. Computations associated with deep models for today's intelligent applications are typically performed solely on the cloud. This cloud-only approach requires significant amounts of raw data to be uploaded to the cloud over the mobile wireless network and imposes considerable computational and communication load on the cloud server. Recent studies have shown that the latency and energy consumption of deep neural networks in mobile applications can be notably reduced by splitting the workload between the mobile device and the cloud. In this approach, referred to as \textit{collaborative intelligence}, intermediate features computed on the mobile device are offloaded to the cloud instead of the raw input data of the network, reducing the size of the data needed to be sent to the cloud. In this paper, we design a new collaborative intelligence friendly architecture by introducing a unit responsible for reducing the size of the feature data needed to be offloaded to the cloud to a greater extent, where this unit is placed after a selected layer of a deep model. This unit is referred to as the \textit{butterfly unit}. The butterfly unit consists of the \textit{reduction unit} and the \textit{ restoration unit.} The outputs of the reduction unit is offloaded to the cloud server on which the computations associated with the restoration unit and the rest of the inference network are performed. Both the reduction and restoration units use a convolutional layer as their main component. The inference outcomes are sent back to the mobile device. The new network architecture, including the introduced butterfly unit after a selected layer of the underlying deep model, is trained end-to-end. Our proposed method, across different wireless networks, achieves on average 53$\times$ improvements for end-to-end latency and 68$\times$ improvements for mobile energy consumption compared to the status quo cloud-only approach for ResNet-50, while the accuracy loss is less than 2\%.
\end{abstract}


\section{Introduction}

Recent advances in deep neural networks (DNNs) have contributed to the state-of-the-art performance in various artificial intelligence (AI)-based applications such as image classification \cite{krizhevsky2012imagenet,Resnet}, object detection \cite{girshick2016region}, speech recognition \cite{hinton2012deep}, natural language processing \cite{mikolov2013distributed}, and so forth. Consequently, mobile and internet of things (IoT) devices are increasingly relying on theses DNNs to improve their performance in such AI-based applications. However, the storage and computation requirements of most of the state-of-the-art deep models limit the fully deployment of the inference network on mobile devices. Therefore, as the most common way for deployment of most of the DNN-based applications on mobile devices, the input data of DNN is sent to cloud servers, and the computations associated with the inference network are performed fully on the cloud side \cite{choi2018deep}.

One of the downsides of the cloud-only approach is the fact that it requires the mobile edge devices to send considerable amounts of data, which can be images, audio, and video, over the wireless network to the cloud. This leads to notable latency and energy overheads on the mobile device. Furthermore, in a scenario where a large number of mobile devices send a vast amount of simultaneous bit streams to the cloud server, the imposed computation workload of simultaneously executing numerous deep models could become a bottleneck on the cloud server.

Recently, inspired by the progress in the computation power and energy efficiency of mobile devices, there has been a body of research studies investigating the strategy of pushing a portion of the workload from cloud servers to mobile edge devices, where both the mobile and cloud execute the inference network collaboratively. As a result, a concept named \textit{collaborative intelligence} has been introduced \cite{eshratifar2018jointdnn,EshratifarGLS,kang2017neurosurgeon,grulich2018collaborative,chen2018intermediate,choi2018near,choi2018deep}. In collaborative intelligence, the deep network is split at an intermediate layer between the mobile and cloud. In other words, instead of sending the original raw data from the mobile device to the cloud and executing the inference network fully on the cloud side, the computations associated with the initial layers are performed on the mobile side. Then, the computed feature tensor of the last assigned layer on the mobile side could be sent to the cloud for executing the remained computation layers of the inference network. By allocating a portion of the inference network to the mobile side, the imposed workload on the cloud reduces, where this results in the increased throughput on the cloud. Furthermore, in some deep models which are based on convolutional neural networks (CNNs), e.g. AlexNet~\cite{krizhevsky2012imagenet}, the feature data volume generally shrinks as we go deeper in the model, and it might become even less than the model input size after a number of layers \cite{kang2017neurosurgeon,eshratifar2018jointdnn,choi2018deep}. Therefore, by computing a few layers on the mobile, the amount of data needed to be sent to the cloud in the collaborative intelligence framework can decrease compared to the cloud-only approach. This can lead to reduced energy and latency overheads on the mobile side.

According to a recent study done in \cite{kang2017neurosurgeon} for different hardware and wireless connectivity configurations, the optimal operating point for the inference network in terms of latency and/or energy consumption is associated with dividing the network between the mobile and cloud, and not the common cloud-only approach, or the mobile-only approach (in case the deep model is able to be executed fully on the mobile device). The optimal point of split depends on the computational and data characterization of DNN layers and is usually at a point deep in the inference network. The approach \cite{eshratifar2018jointdnn} has extended the work of \cite{kang2017neurosurgeon} and included model training and additional network architectures. The network is again split between the mobile and cloud, but the data can flow in both ways in order to optimize the efficiency of both the inference and training. 
In summary, in the research works studying the collaborative intelligence framework, a given deep network is split between the mobile device and the cloud without any modification to the network architecture itself \cite{eshratifar2018jointdnn,EshratifarGLS,kang2017neurosurgeon,grulich2018collaborative,chen2018intermediate,choi2018near,choi2018deep}. 

\begin{figure}
\begin{center}
  \includegraphics{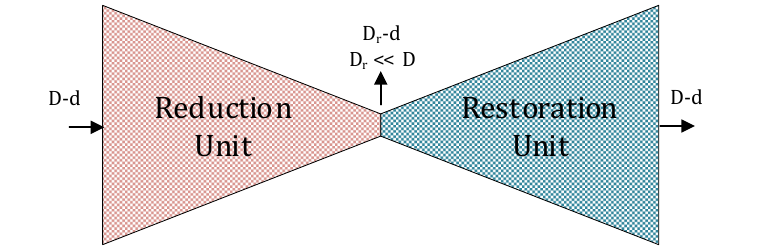}
  \caption{The butterfly unit. It takes a tensor with $D$ channels and shrinks it into a tensor with $D_r$ channels using the reduction unit, where $D_r \ll D$. It outputs a tensor with the same dimension as input using the restoration unit.}
  \label{fig:buterfly}
\end{center}
\end{figure}

In this paper, we investigate the problem of altering a given deep network architecture before the partitioning of it between the mobile and cloud. For this purpose, on the mobile side, we introduce a \textit{reduction unit} right before uploading the feature tensor to the cloud. The reduction unit is stacked to the end of the computation layers assigned to the mobile side. The computation associated with the reduction unit is also done on the mobile side. The purpose of this unit is reducing the feature data volume needed to be sent to the cloud via the wireless network to a greater extent, since the latency and energy overheads associated with the wireless upload link in state-of-the-art approaches for collaborative intelligence still contribute to the major portion of the energy consumption of the inference network on the mobile side and end-to-end latency \cite{eshratifar2018jointdnn}. Accordingly, on the cloud side, we introduce a \textit{restoration unit} which is stacked before the computation layers assigned to the cloud. Both the reduction and restoration units use a convolutional layer as their main component. The dimension of the convolution layers used in reduction and restoration units are set in a way so that the dimension of the input tensor of the reduction unit is equal to the dimension of the output tensor of the restoration unit. We refer to the combination of reduction and restoration units as the \textit{butterfly unit} (see Fig. \ref{fig:buterfly}).
The new network architecture, including the introduced butterfly unit after a selected layer of the underlying deep model, is trained end-to-end, while in other works which have considered compression for reducing the feature data volume needed to be sent to the cloud, they have added non-learnable compression techniques (e.g. JPEG) to an already trained model \cite{choi2018near,choi2018deep,chen2018intermediate}. 

The rest of the paper is structured as follows: Section \ref{section.proposed_method} elaborates more on the details of the proposed butterfly unit and the proposed DNN partitioning algorithm. Section \ref{section.evaluation} provides the obtained improvements in terms of end-to-end latency and the mobile energy consumption. It also discusses the flexibility of network partitioning point based on the load level of the cloud and mobile, and the wireless network conditions. Finally, Section \ref{section.conclusion} concludes the paper.

\section{Proposed Method} \label{section.proposed_method}
\subsection{Butterfly Unit} \label{section.butterfly}
\begin{figure}
\begin{center}
  \includegraphics[width=\columnwidth]{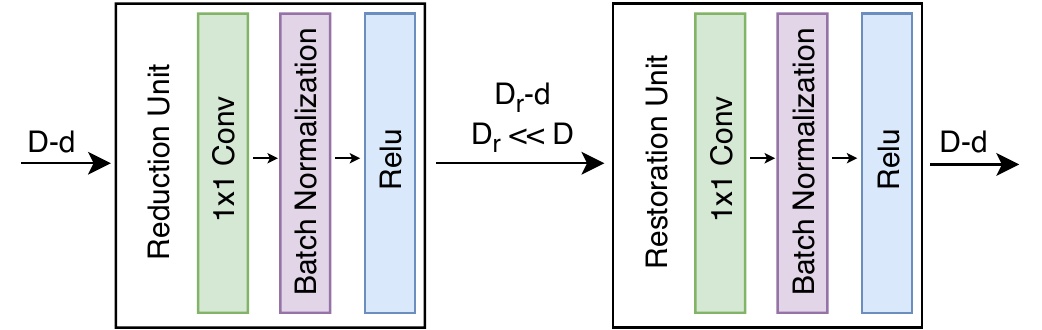}
  \caption{The butterfly unit architecture. It consists of the reduction unit on the mobile side, and the restoration unit on the cloud side.}
  \label{fig:bottleneck_unit}
\end{center}
\end{figure}

The butterfly unit, as shown in Fig.~\ref{fig:bottleneck_unit}, consists of two components: 1) the reduction unit, and 2) the restoration unit. The input to the reduction unit is a tensor of size ($batch\_size$, $width$, $height$, $D$). A convolution filter of size ($1$, $1$, $D$, $D_r$) is applied to the input, producing a tensor of size ($batch\_size$, $width$, $height$, $D_r$) as the output of the reduction unit. The output tensor of the reduction unit is the shrunk representation of its input along the channel axis ($D_r$ $\ll$ $D$), and it is the tensor which is uploaded to the cloud server. On the cloud side, in the restoration unit, by applying a convolution filter of size ($1$, $1$, $D_r$, $D$), we restore the dimension of the original input of the butterfly unit to proceed the rest of the inference. 
The butterfly unit is placed after one of the layers in a DNN. The intuition behind decreasing the tensor dimension along the channel axis in the reduction unit is the fact that typically each channel preserves the visual structure of the input. Therefore, we can expect this non-expensive 1$\times$1 convolution can keep enough information of the feature data. In addition, depending on the architecture of the underlying deep model, the feature tensor size varies layer by layer, typically increasing in channel sizes. 
Therefore, as we go deeper in the model, more channels would be required in the output of the reduction unit, $D_r$, for maintaining the accuracy of the model. 

From the perspective of the mobile device, the location of the butterfly unit is desired to be closer to the input layer so that the mobile device computes fewer layers. However, from the perspective of the cloud server, we want to push more computations towards the mobile device in order to reduce the data center workloads. Particularly, when the cloud server and/or the wireless network are congested, pushing computations towards the mobile device is advantageous. As a result, there is a trade-off in choosing the location of the butterfly unit in the inference network. 

\subsection{Partitioning Algorithm}
The proposed algorithm, for choosing the location of the butterfly unit and the proper value of $D_r$, comprises three main steps: 1) Training, 2) Profiling, and 3) Selection. In the training phase, we train $M$ models, where each model is associated with placing the butterfly unit after a different arbitrary layer among the total $N$ layers of the inference network ($M \leq N$). For each model, via linear search, we find and choose the minimum $D_r$ for the butterfly unit that reaches a pre-defined acceptable accuracy. During the profiling phase, for each of $M$ models, we measure the latency values corresponding to the computation of layers assigned to the mobile side, the reduction unit, the wireless up-link of the shrunk feature data to the cloud, the restoration unit, and the computations of layers assigned to the cloud side. Furthermore, for energy consumption, we measure the values associated with the computation of layers assigned to the mobile side, the reduction unit, and the wireless up-link of the shrunk feature data. These measurements vary for each of $M$ models, and the current load level of the mobile, cloud, or wireless network conditions. In the end, in the selection phase, depending on whether the target is minimizing the end-to-end latency or the mobile energy consumption, we select the best partitioning among $M$ available options.

The full procedure for choosing the location of the butterfly unit and the proper value of $D_r$ is shown in Algorithm~\ref{the_algorithm}.

\begin{algorithm}
    \SetKwInOut{Input}{Input}
    \SetKwInOut{Output}{Output}

\textbf{Inputs:}\\
$N$: number of layers in the DNN\\
$M$: number of partitioning points in the DNN ($M \leq N$)\\
$\{P_j|j=1..M\}$: location of each partition point\\
$\{F_i|i=1..N\}$: feature data size at each layer\\
$\{C_i|i=1..N\}$: output channel size of each layer\\
$K_{mobile}$: current load level of mobile\\
$K_{cloud}$: current load level of cloud\\
$t_{mobile}, p_{mobile}(j,K_{mobile}) |j=1..M$: latency and power on the mobile corresponding to partition $j$ and load $K_m$\\
$t_{cloud}(j,K_{cloud}) |j=1..M$: latency on the cloud corresponding to partition $j$ and load $K_{cloud}$\\
$NB$: wireless network bandwidth\\
$PU$: wireless network up-link power consumption\\
\textbf{Output:} Best partitioned model
\\
~\\
// Training phase \\
 \For{$j = 1;\ j \leq M;\ j = j + 1$}{
    \For{$k = 1;\ k \leq C_{P_j};\ k = k + 1$} {
        Place butterfly unit of $D_r$ = $k$ after $P_j$\\
        Train()\\
        \If{accuracy is acceptable}{
            Store as $j$th partitioned model\\
            break
            }
    }
 }
 ~\\
 // Profiling phase \\
  \For{$j = 1;\ j \leq M;\ j = j + 1$}{
    $TM_j$ = $t_{mobile}(j, K_{mobile})$\\
    $PM_j$ = $p_{mobile}(j, K_{mobile})$\\
    $TC_j$ = $t_{cloud}(j, K_{cloud})$\\
    $TU_j$ = $F_{P_j}/NB$
    }
    ~\\
// Selection phase \\
     \If{target is min latency} {
    	return $argmin_{j=1..M} (TM_j + TU_j + TC_j)$
      }
      \If{target is min energy} {
        return $argmin_{j=1..M} (TM_j \times PM_j + TU_j \times PU)$
      }
    \caption{The proposed DNN partitioning algorithm}
    \label{the_algorithm}
\end{algorithm}

\begin{figure}
\begin{center}
  \includegraphics{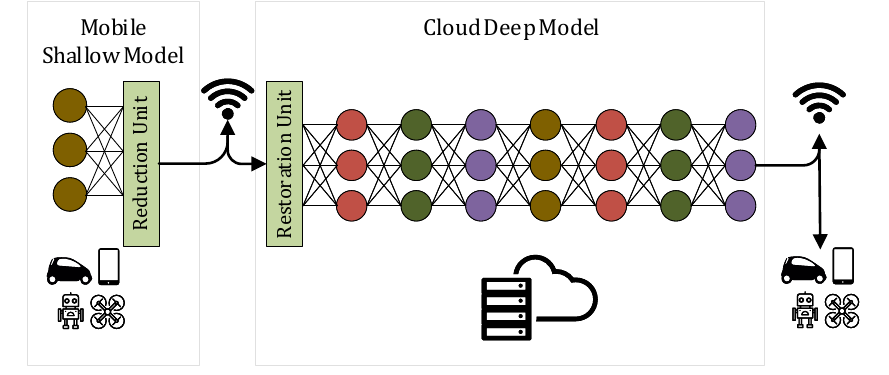}
  \caption{The proposed method overview. A shallow model and the reduction unit on the mobile device extracts a dense representation of the input, which is uploaded to the cloud. Then, on the cloud, after applying the restoration function on the dense representation, the rest of the inference procedure is followed. }
  \label{fig:overall_architecture}
\end{center}
\end{figure}

\section{Evaluation} \label{section.evaluation}


\begin{figure*}
  \begin{center}
  \includegraphics[width=\textwidth]{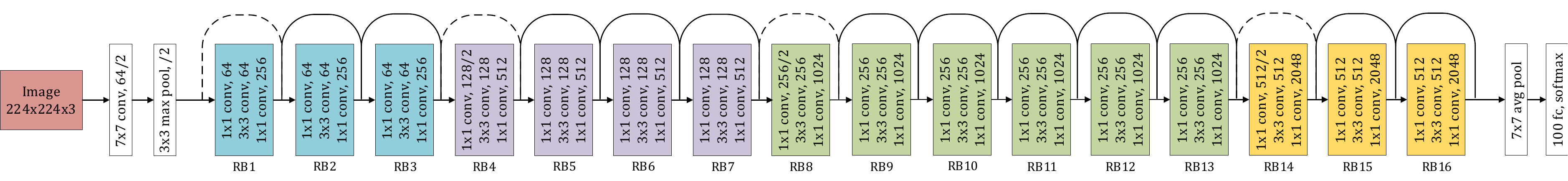}
  \caption{ResNet-50 architecture and its 16 residual blocks. The solid lines represent identity shortcuts, and the dashed lines represent projection shortcuts. }
  \label{fig:resnet50}
  \end{center}
\end{figure*}

\begin{figure}
\begin{center}
  \includegraphics[width=\columnwidth]{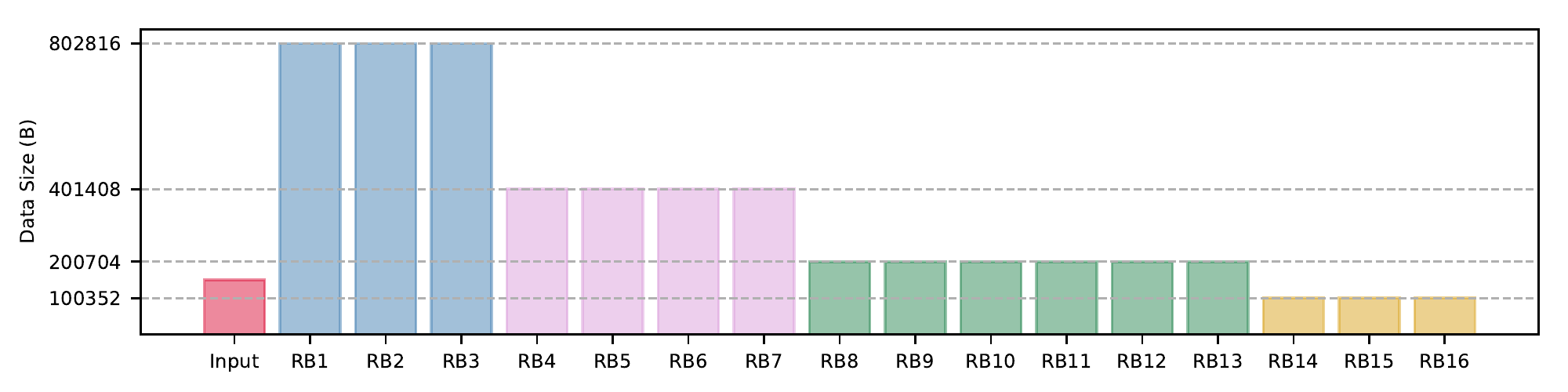}
  \caption{Input image size of the model and the size of output feature tensor of each residual block in ResNet-50.}
  \label{fig:resnet_feature_size}
\end{center}
\end{figure}

\begin{figure}
\begin{center}
  \includegraphics{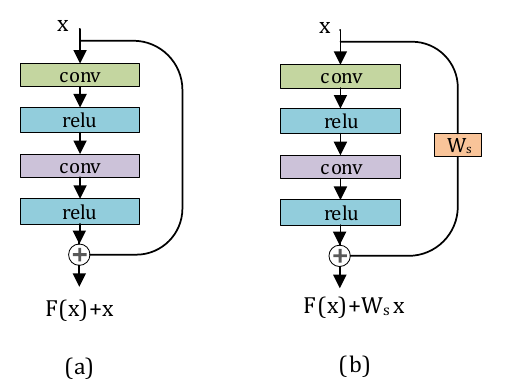}
  \caption{Residual block architecture with (a) identity and (b) projection shortcut. }
  \label{fig:residual_block}
\end{center}
\end{figure}

\subsection{Experimental Setup}\label{AA}
We evaluate our proposed method on NVIDIA Jetson TX2 board\cite{JetsonTX2}, which is equipped with NVIDIA Pascal\texttrademark\space GPU which fairly represents the computing power of mobile devices. Our server platform is equipped with NVIDIA Geforce\textregistered\space GTX 1080 Ti GPU, which has almost 30x more computing power compared to our mobile platform. The detailed specifications of our mobile and server platforms are presented in Table \ref{table:mobile_platform} and Table \ref{table:server_platform}, respectively. We measure the GPU power consumption on our mobile platform using INA226 power monitoring sensor with a sampling rate of 500~kHz\cite{INA226}. 
For our wireless network settings, the average upload speed of different wireless networks, 3G, 4G, and Wi-Fi, in the U.S. are used in our experiments~\cite{MobNet,Speedtest}. We use the transmission power models of \cite{4GLTE} for wireless networks, which have estimation errors less than 6\%. The power level for up-link is $P_u = \alpha_u t_u + \beta$ 
, where $t_u$ is the up-link throughput, and $\alpha_u$ and $\beta$ are regression coefficients of power models. The values for our power model parameters are presented in Table~\ref{table:network_parameters}.

We prototype the proposed method by implementing the inference networks, both for the mobile device and cloud server, using NVIDIA TensorRT\texttrademark\space\cite{TensorRT}, which is a platform for high-performance deep learning inference. It includes a deep learning inference optimizer and run-time that delivers low latency and high-throughput for deep learning inference applications. TensorRT is equipped with cuDNN\cite{cuDNN}, a GPU-accelerated library of primitives for DNNs. TensorRT supports three precision modes for creating the inference graph: FP32 (single precision), FP16 (half precision), and INT8. However, our mobile device does not support INT8 operations on its GPU. Therefore, we use FP16 mode for creating the inference graph from the trained model graph, where for training itself single precision mode (32-bit) is used. As demonstrated in \cite{deepcompression}, 8-bit quantization would be enough for even challenging tasks like ImageNet~\cite{ImageNet} classification. Therefore, we quantize FP16 data types to 8 bits only for uploading the feature tensor to the cloud. We implement our client-server interface using Thrift~\cite{slee2007thrift}, an open source flexible RPC interface for inter-process communication. To allow for flexibility in the dynamic selection of partition points, both the mobile and cloud host all possible $M$ partitioned models. For each of $M$ models, the mobile and cloud store only their assigned layers. At run-time, depending on the load of the mobile and cloud, wireless network conditions, and the optimization goal (minimizing for latency or energy), only one of $M$ partitioned models is selected. Given a partition decision, execution begins on the mobile device and cascades through the layers
of the DNN leading up to that partition point. Upon completion of that layer and the reduction unit, the mobile sends the output of the reduction unit from the mobile device to the cloud. Cloud server then executes the computations associated with the restoration unit and remaining DNN layers. Upon the completion of the execution of the last DNN layer on the cloud, the final result is sent back to the mobile device. 

We evaluate the proposed method on one of the promising and mostly used DNN architectures, ResNet~\cite{Resnet}. DNNs are hard to train because of the notorious vanishing/exploding gradient issue, which hampers the convergence of the model. As a result, as the network goes deeper, its performance gets saturated or even starts degrading rapidly~\cite{jousha_difficulty}. The core idea of ResNet is introducing a so-called “identity shortcut connection” that skips one or more layers. The output of a residual block (RB) with identity mapping will be $y = F(x,W) + x$ instead of traditional $y = F(x,W)$. The argument behind ResNet's good performance is that stacking layers should not degrade the network performance, because we could simply stack identity mappings (layers that do nothing, i.e., $y = x$) upon the current model, and the resulting architecture would perform the same. It indicates that the deeper model should not produce a training error higher than its shallower counterparts. They hypothesize that letting the stacked layers fit a residual mapping is easier than letting them directly fit the desired underlying mapping. 
If the dimensions change, there are two cases: 1) increasing dimensions: The shortcut still performs identity mapping, with extra zero entries padded with the increased dimension. 2) decreasing dimensions: A projection shortcut is used to match the dimensions of $x$ and $y$ using the formula of $y = F(x,W) + W_s x$, as shown in Fig.~\ref{fig:residual_block}.

ResNet architecture comes with flexible number of layers (e.g. 34, 50, 101, etc.).
In our experiments, we use ResNet-50. There are 16 residual blocks in ResNet-50 ~\cite{Resnet}. Using Algorithm \ref{the_algorithm}, we obtain 16 models where each model is corresponding to placing the butterfly unit after one of 16 residual blocks. The detailed architecture, and the data size of layer outputs of ResNet-50 are demonstrated in Fig.~\ref{fig:resnet50}, and Fig.~\ref{fig:resnet_feature_size}, respectively. As indicated in Fig.~\ref{fig:resnet_feature_size}, the size of intermediate feature tensors in ResNet-50 are larger than the input size up until RB14, which is relatively deep in the model. Therefore, merely splitting this network between the mobile and cloud for collaborative intelligence may not perform better than the cloud-only approach in terms of latency and mobile energy consumption, since a large portion of the workload is pushed toward the mobile.

We evaluate the proposed method on miniImageNet~\cite{miniImageNet} dataset, a subset of ImageNet dataset, which includes 100 classes and 600 examples per each class. We use 85\% of whole dataset examples as the training set and the rest as the test set. We randomly crop a 224$\times$224 region out of each sample for data augmentation and train each of the models for 90 epochs.


\begin{table}[ht]
\caption{Mobile device specifications} 
\centering 
\begin{tabular}{|c|c|} 
\hline 
\textbf{Component}& \textbf{Specification} \\ [0.5ex] 
\hline 
System & NVIDIA Jetson TX2 Developer Kit \\
\hline
GPU & NVIDIA Pascal\texttrademark, 256 CUDA cores \\
\hline
CPU & HMP Dual Denver + Quad ARM\textregistered\space A57/2 MB L2 \\
\hline
Memory & 8 GB 128 bit LPDDR4 59.7 GB/s \\ 
\hline 
\end{tabular}
\label{table:mobile_platform} 
\end{table}

\begin{table}[ht]
\caption{Server platform specifications} 
\centering 
\begin{tabular}{|c|c|} 
\hline 
\textbf{Component}& \textbf{Specification} \\ [0.5ex] 
\hline 
GPU & NVIDIA Geforce\textregistered\space GTX 1080 Ti, 12GB GDDR5\\
\hline
CPU & Intel\textregistered\space Xeon\textregistered\space CPU E7- 8837  @ 2.67GHz \\
\hline
Memory & 64 GB DDR4\\ 
\hline 
\end{tabular}
\label{table:server_platform} 
\end{table}

\begin{table}[h]
	\caption{\textcolor{black}{Wireless networks parameters}} 
	\centering 
	\begin{tabular}{|c|c|c|c|} \hline
	\textbf{Param.} & \textbf{3G} & \textbf{4G} & \textbf{Wi-Fi} \\ \hline
	$t_u$ (Mbps)	&	1.1		&	5.85	&	18.88 \\ \hline
$\alpha_u$ (mW/Mbps)	& 868.98	& 438.39	&	283.17    \\ \hline
$\beta$ (mW)	& 817.88	& 1288.04	&	132.86            \\ \hline
\end{tabular}
\label{table:network_parameters} 
\end{table}

\begin{figure*}[!htb]
\minipage{0.124\textwidth}
  \includegraphics[width=\linewidth]{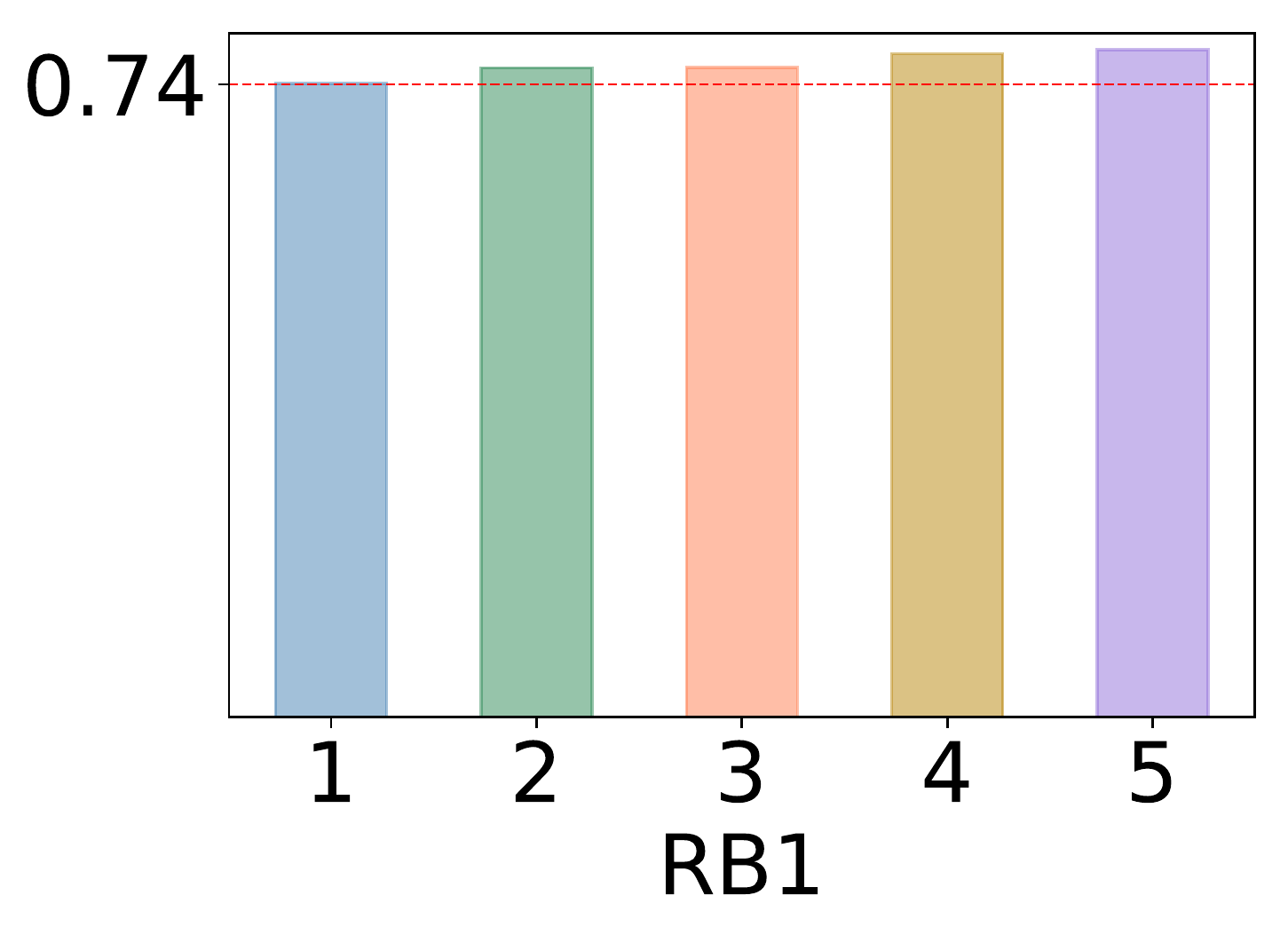}
\endminipage\hfill
\minipage{0.124\textwidth}
  \includegraphics[width=\linewidth]{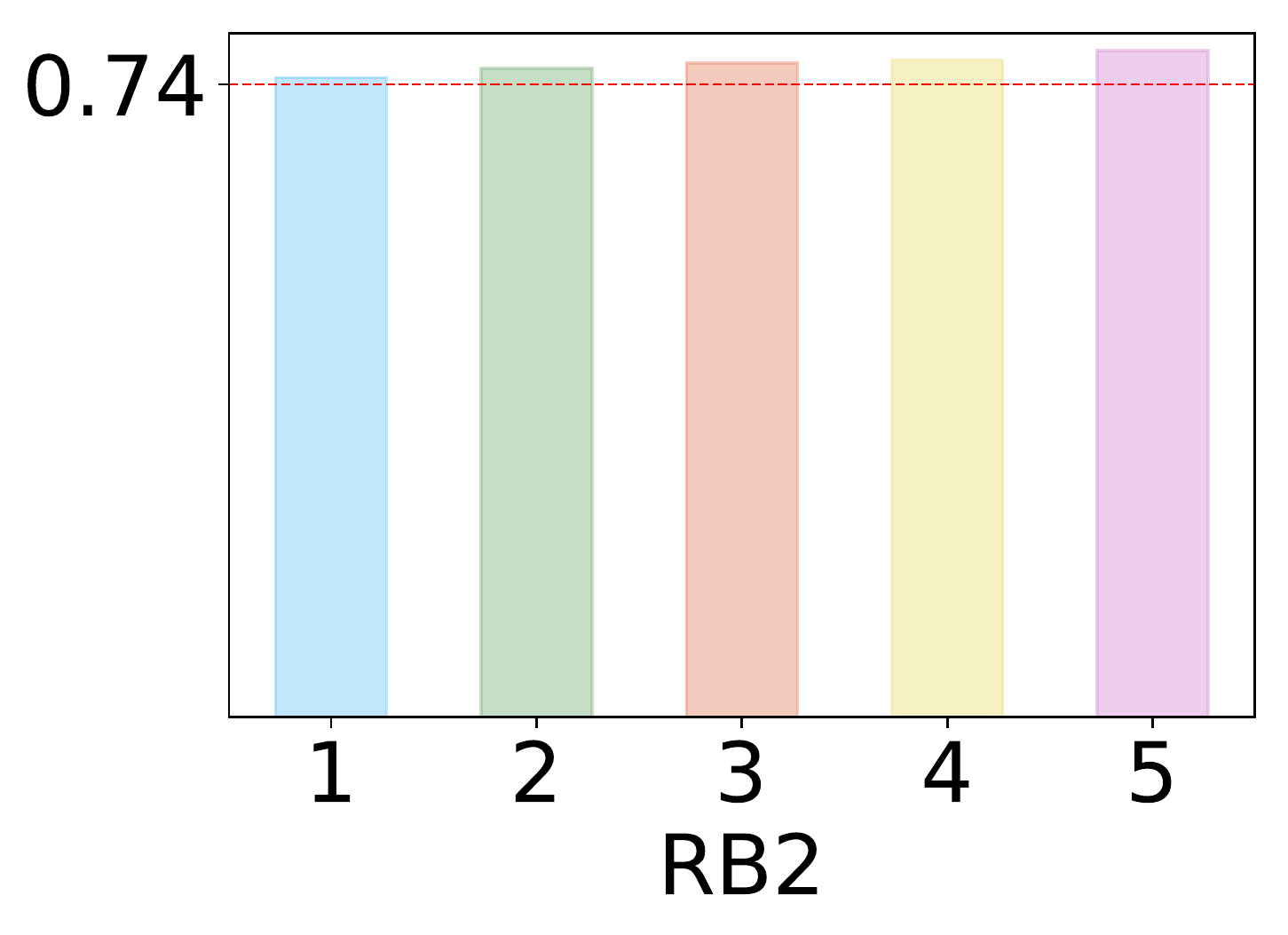}
\endminipage\hfill
\minipage{0.124\textwidth}%
  \includegraphics[width=\linewidth]{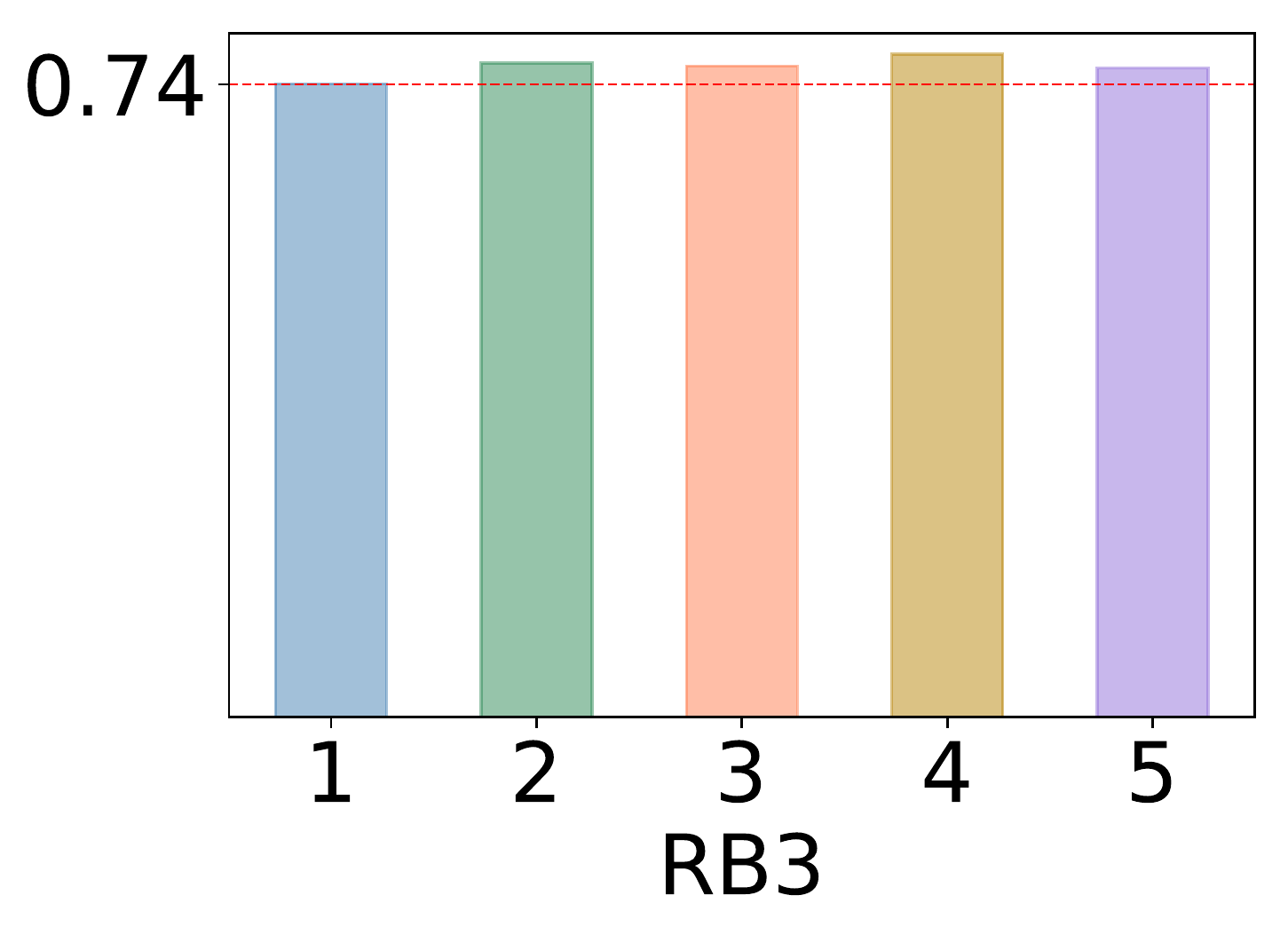}
  
\endminipage\hfill
\minipage{0.124\textwidth}%
  \includegraphics[width=\linewidth]{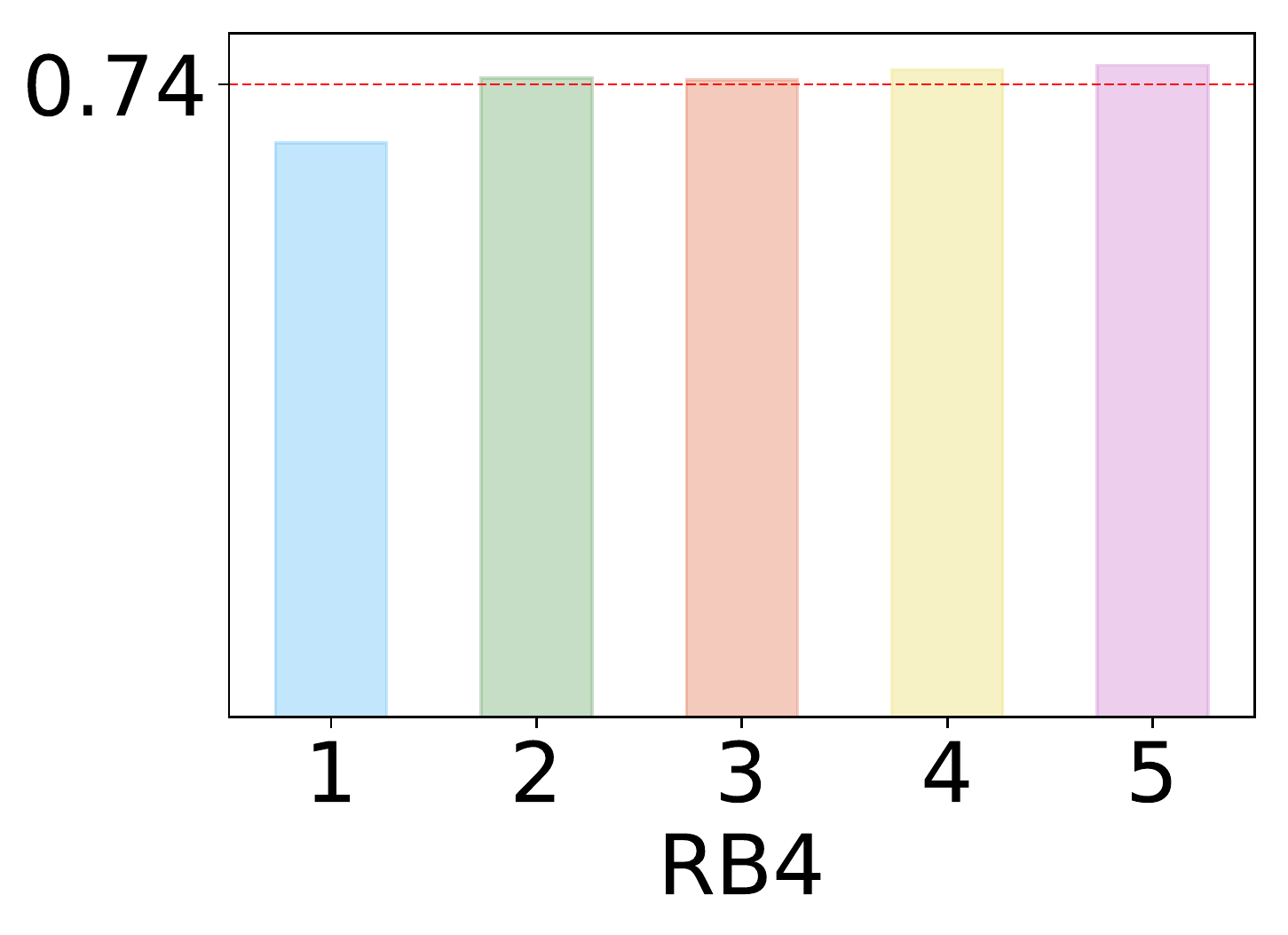}
  
\endminipage\hfill
\minipage{0.124\textwidth}%
  \includegraphics[width=\linewidth]{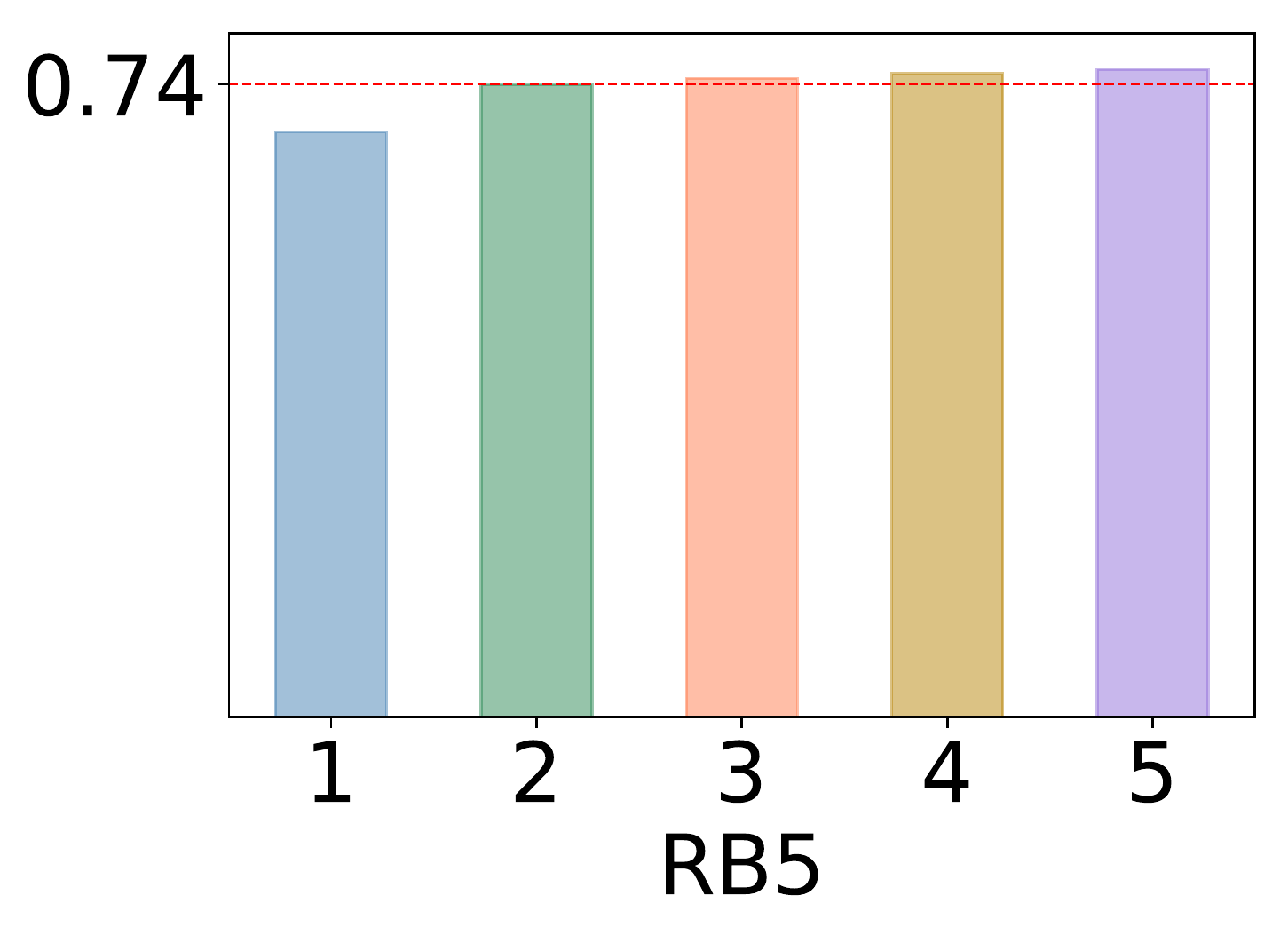}
  
\endminipage\hfill
\minipage{0.124\textwidth}%
  \includegraphics[width=\linewidth]{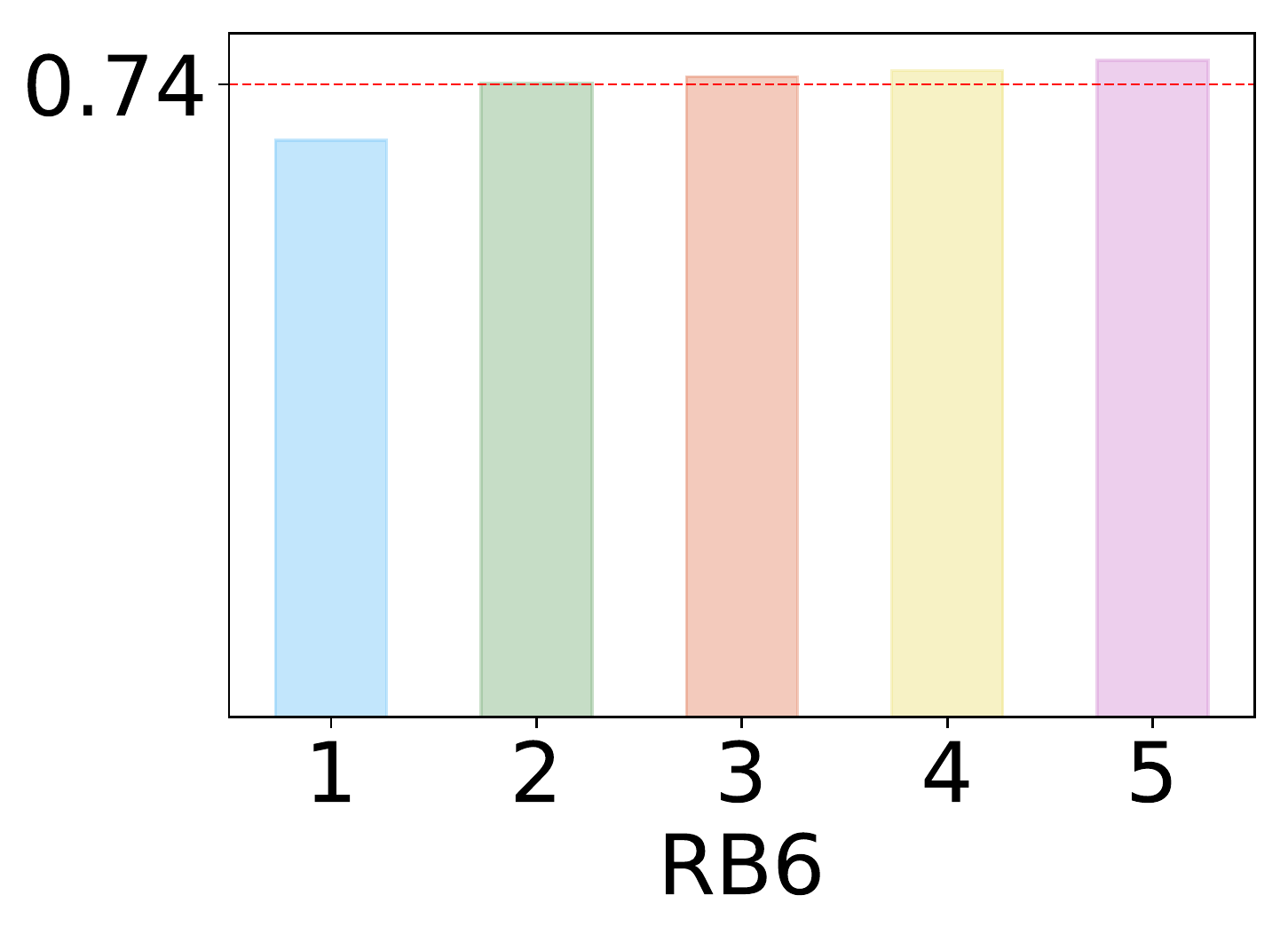}
  
\endminipage\hfill
\minipage{0.124\textwidth}%
  \includegraphics[width=\linewidth]{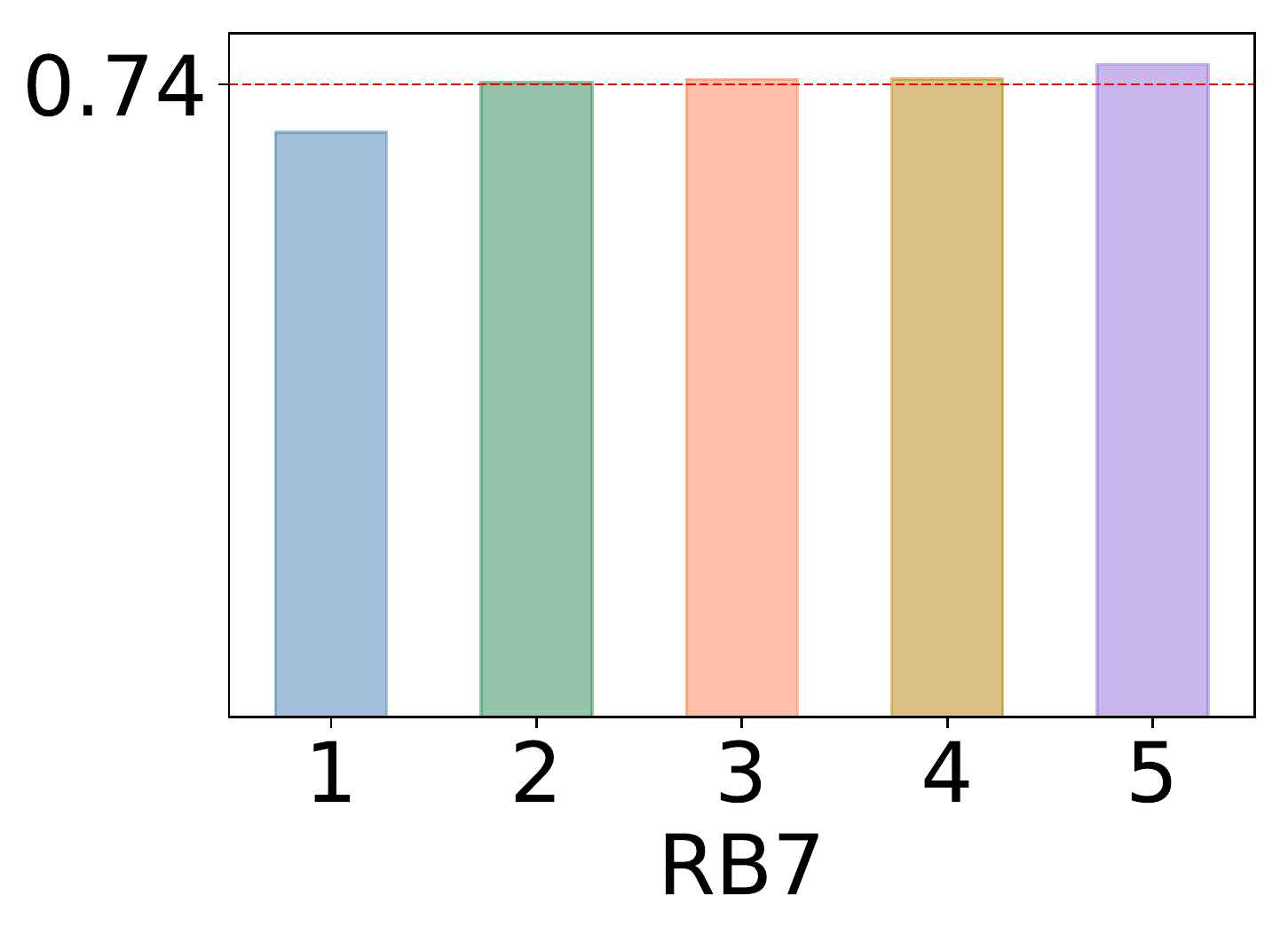}
  
\endminipage\hfill
\minipage{0.124\textwidth}%
  \includegraphics[width=\linewidth]{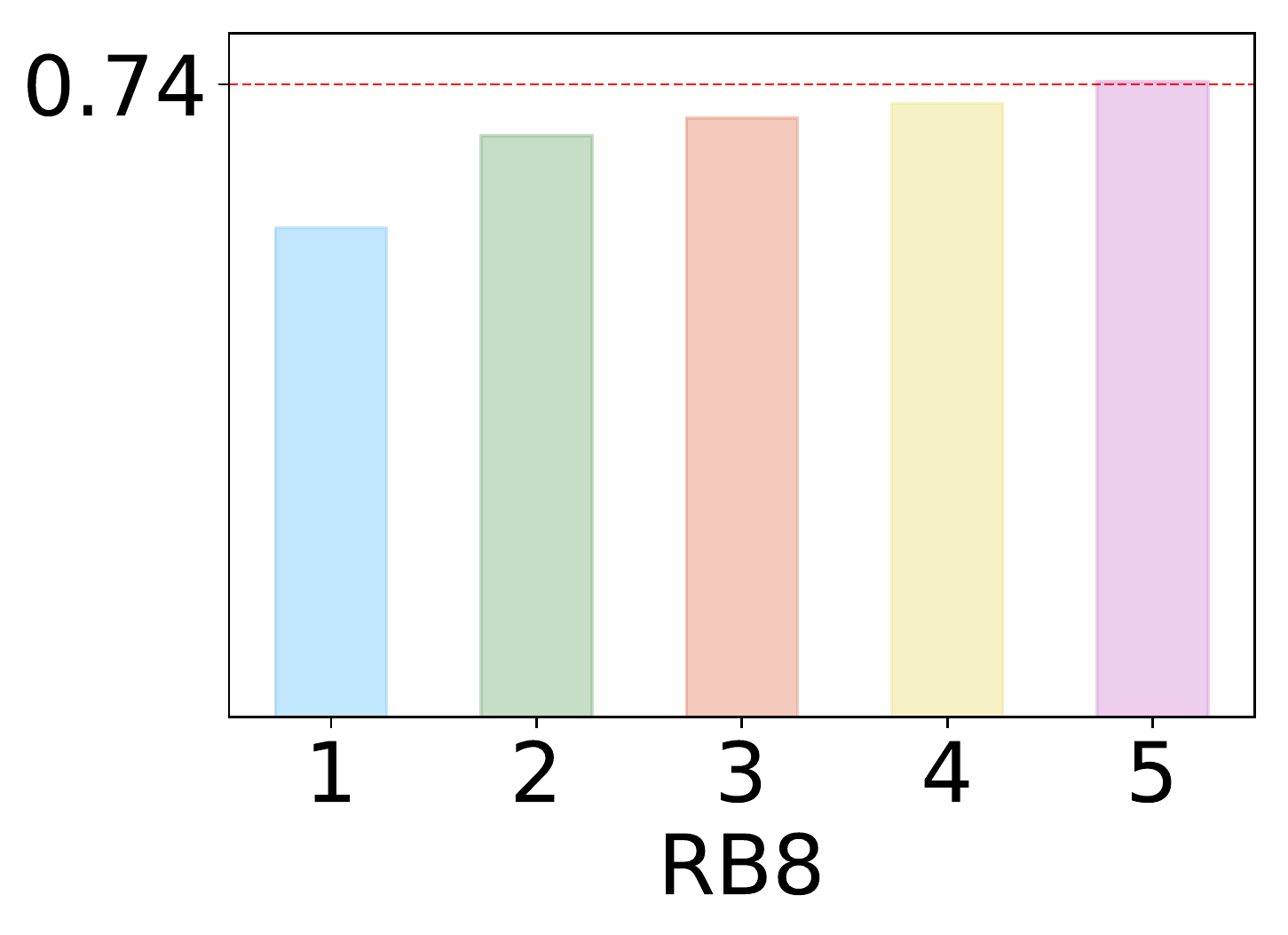}
  
\endminipage\hfill

\minipage{0.124\textwidth}
  \includegraphics[width=\linewidth]{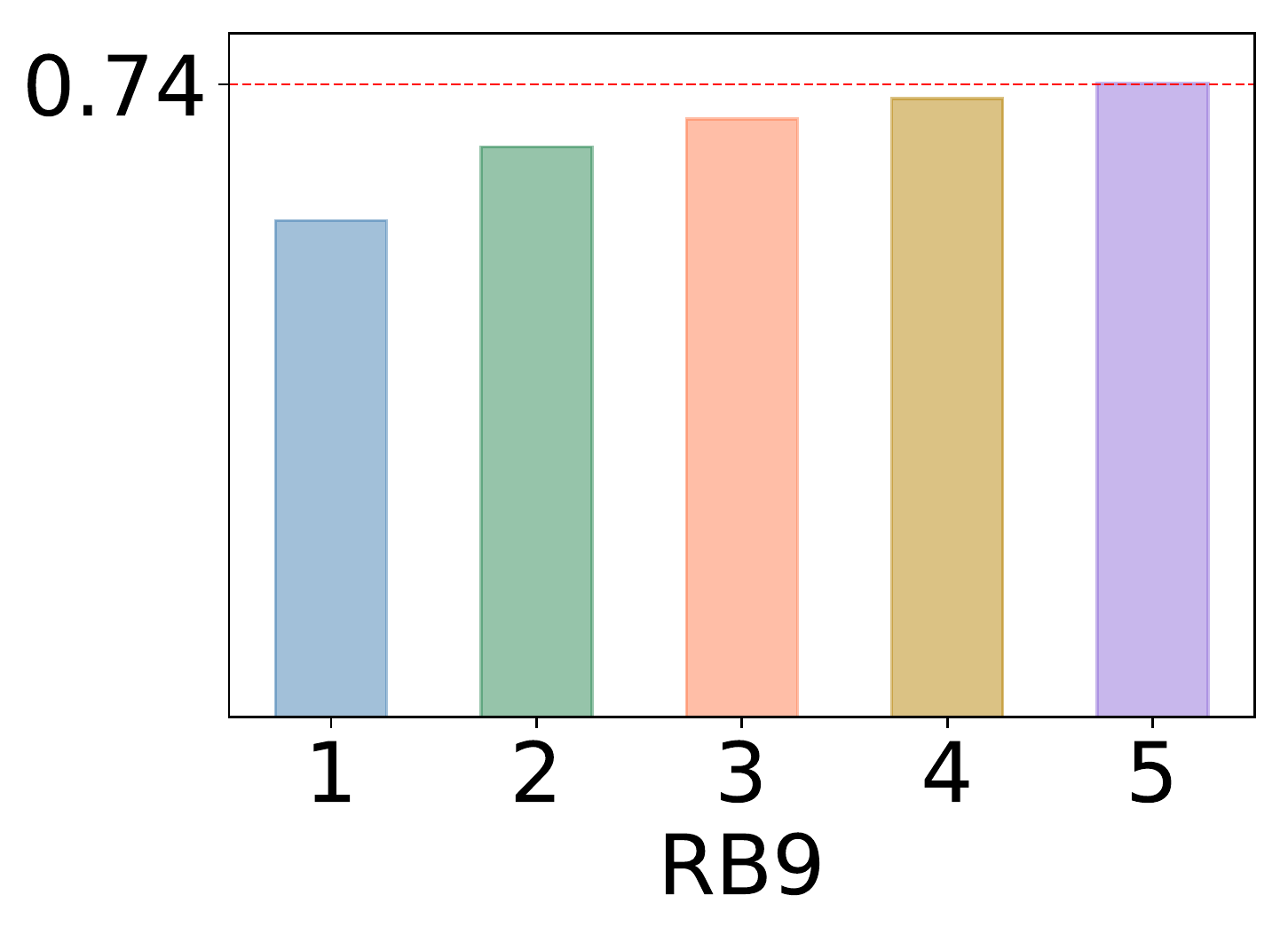}
\endminipage\hfill
\minipage{0.124\textwidth}
  \includegraphics[width=\linewidth]{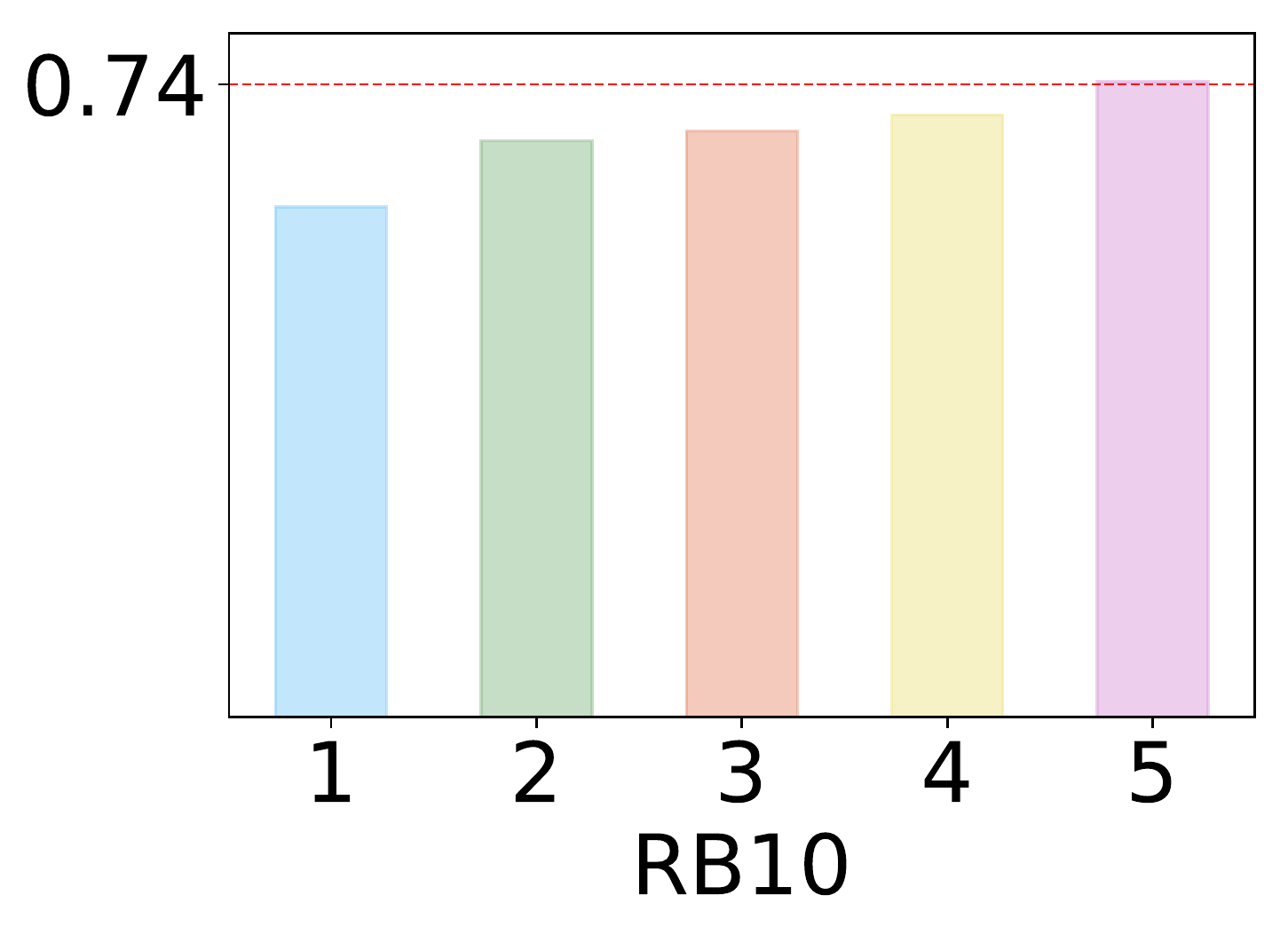}
\endminipage\hfill
\minipage{0.124\textwidth}%
  \includegraphics[width=\linewidth]{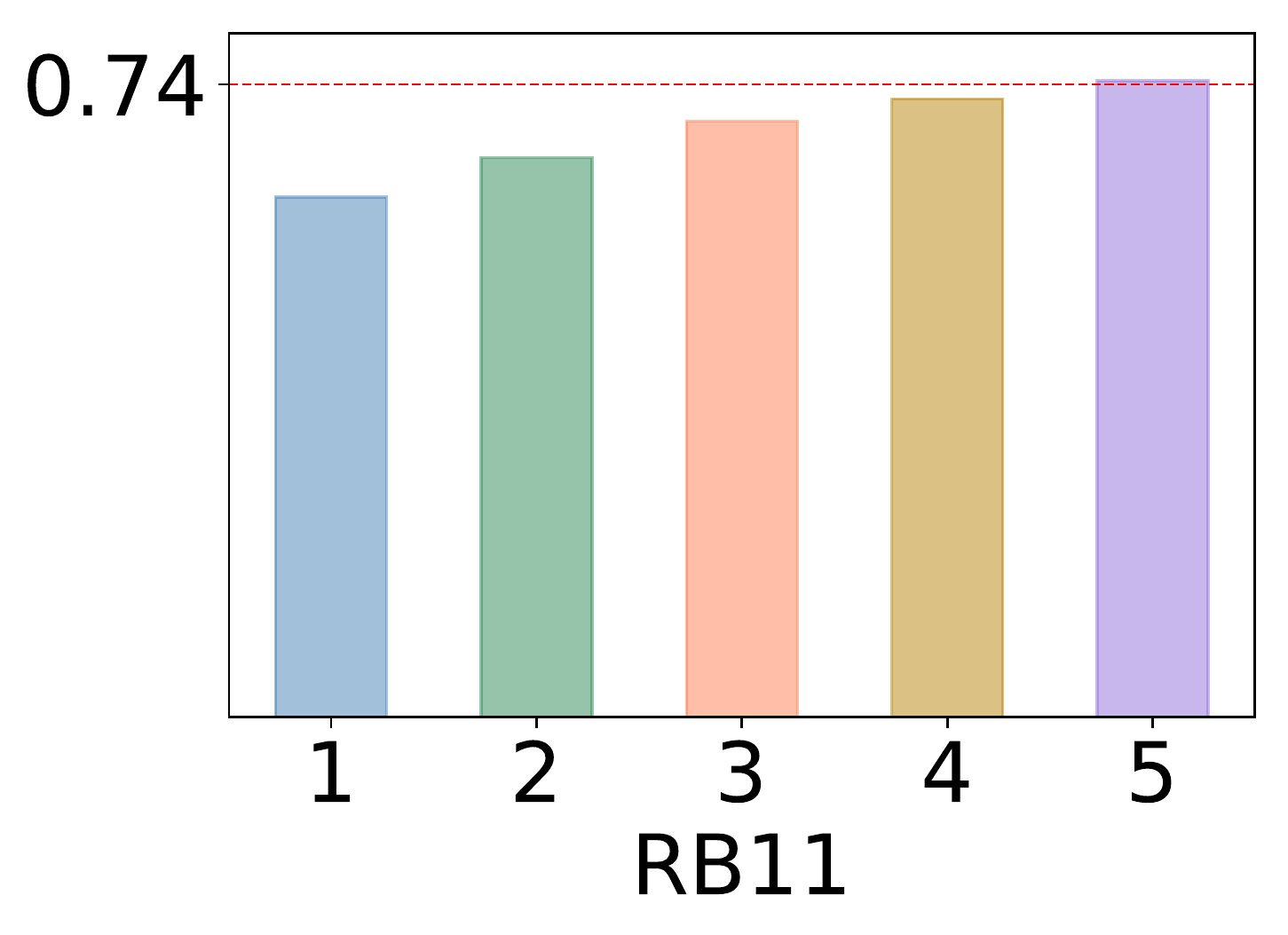}
  
\endminipage\hfill
\minipage{0.124\textwidth}%
  \includegraphics[width=\linewidth]{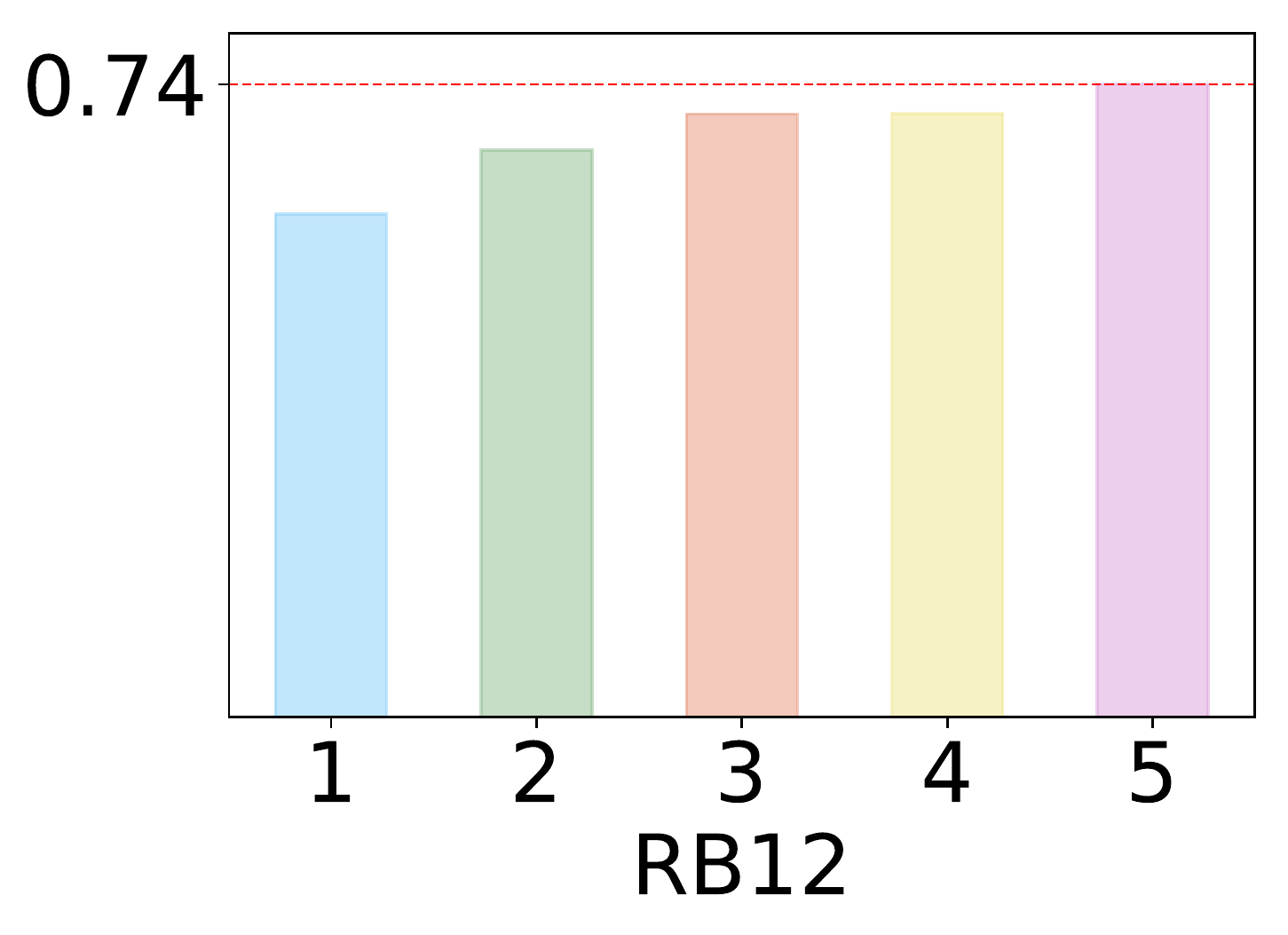}
  
\endminipage\hfill
\minipage{0.124\textwidth}%
  \includegraphics[width=\linewidth]{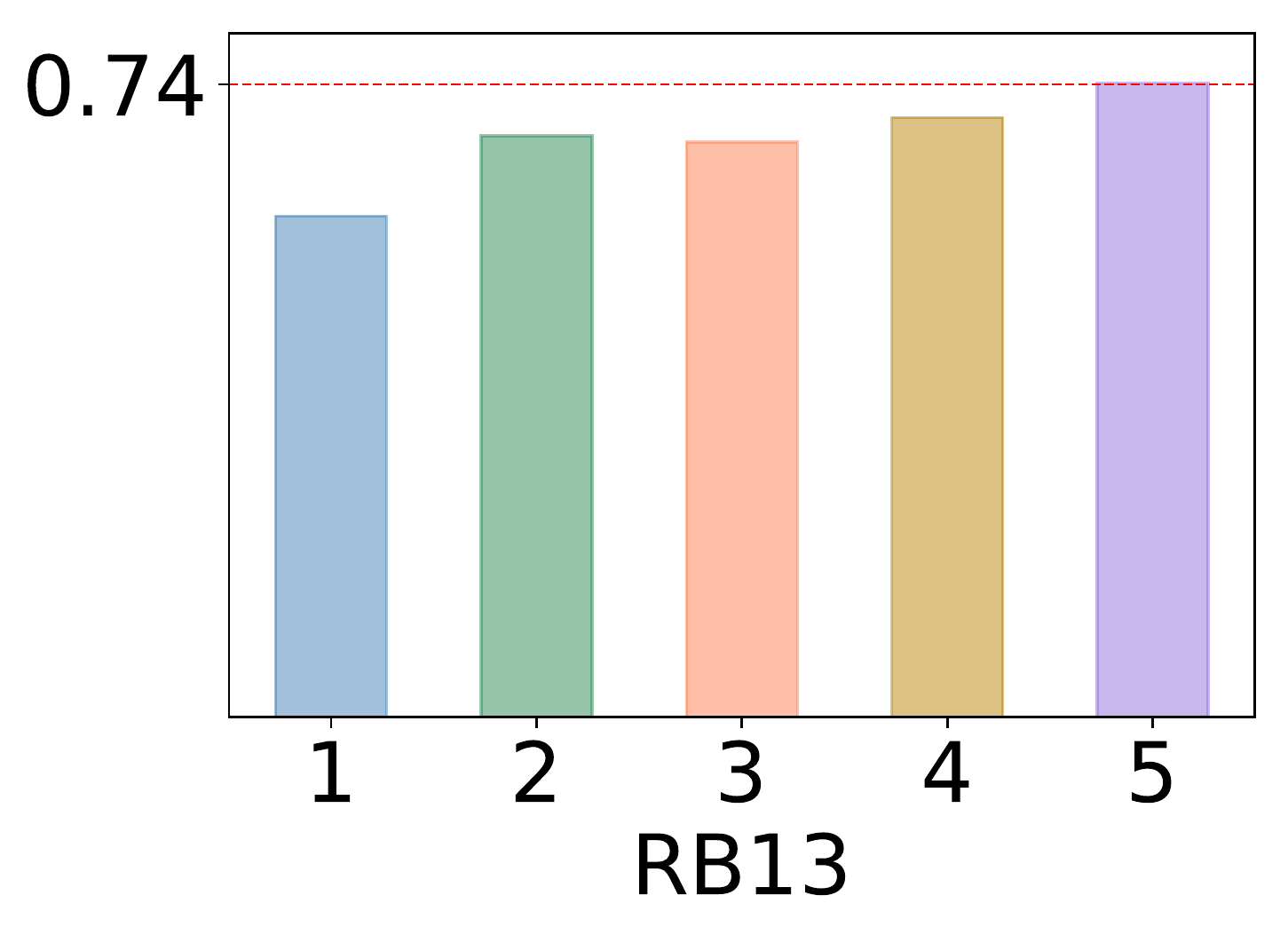}
  
\endminipage\hfill
\minipage{0.124\textwidth}%
  \includegraphics[width=\linewidth]{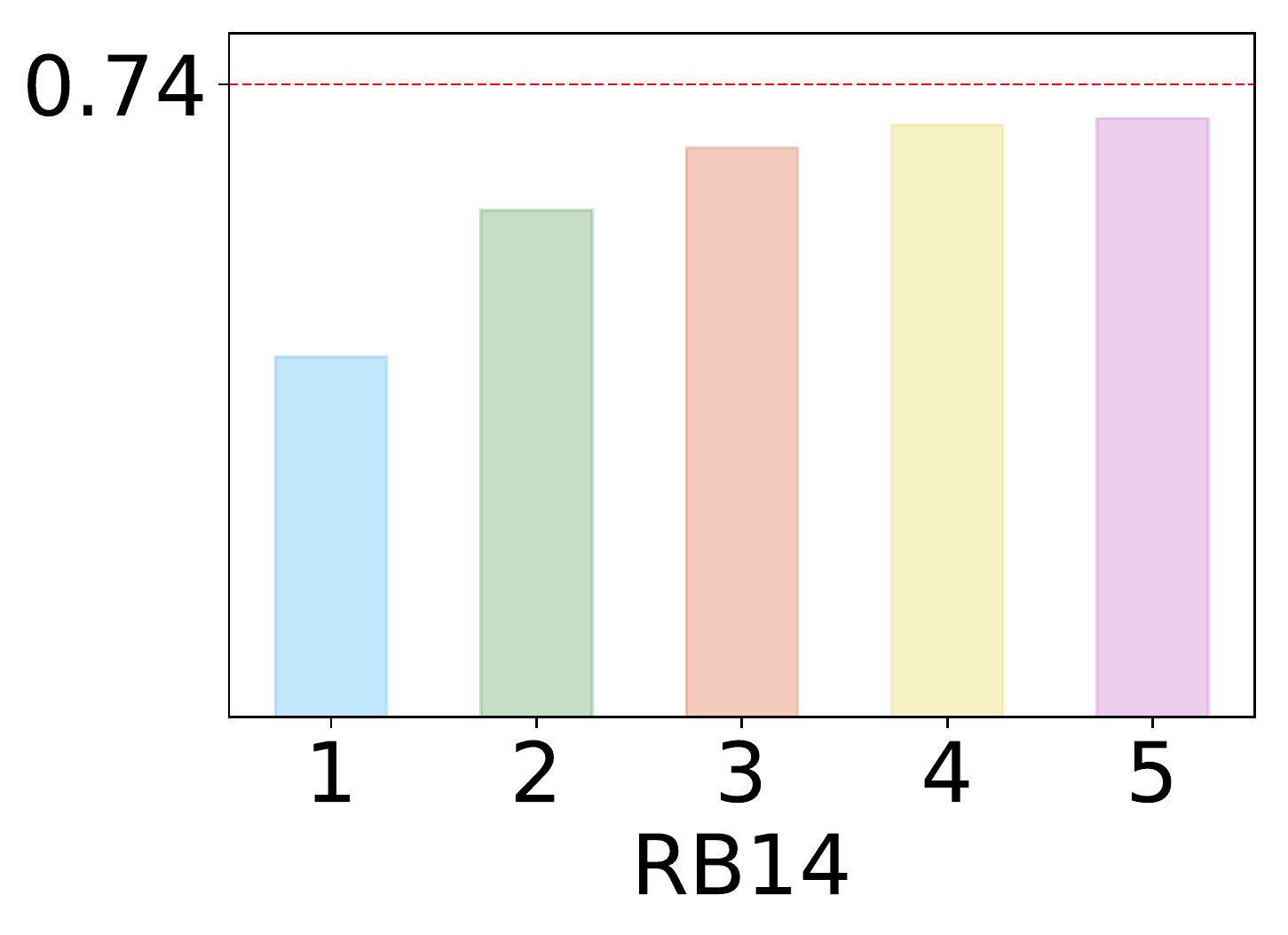}
  
\endminipage\hfill
\minipage{0.124\textwidth}%
  \includegraphics[width=\linewidth]{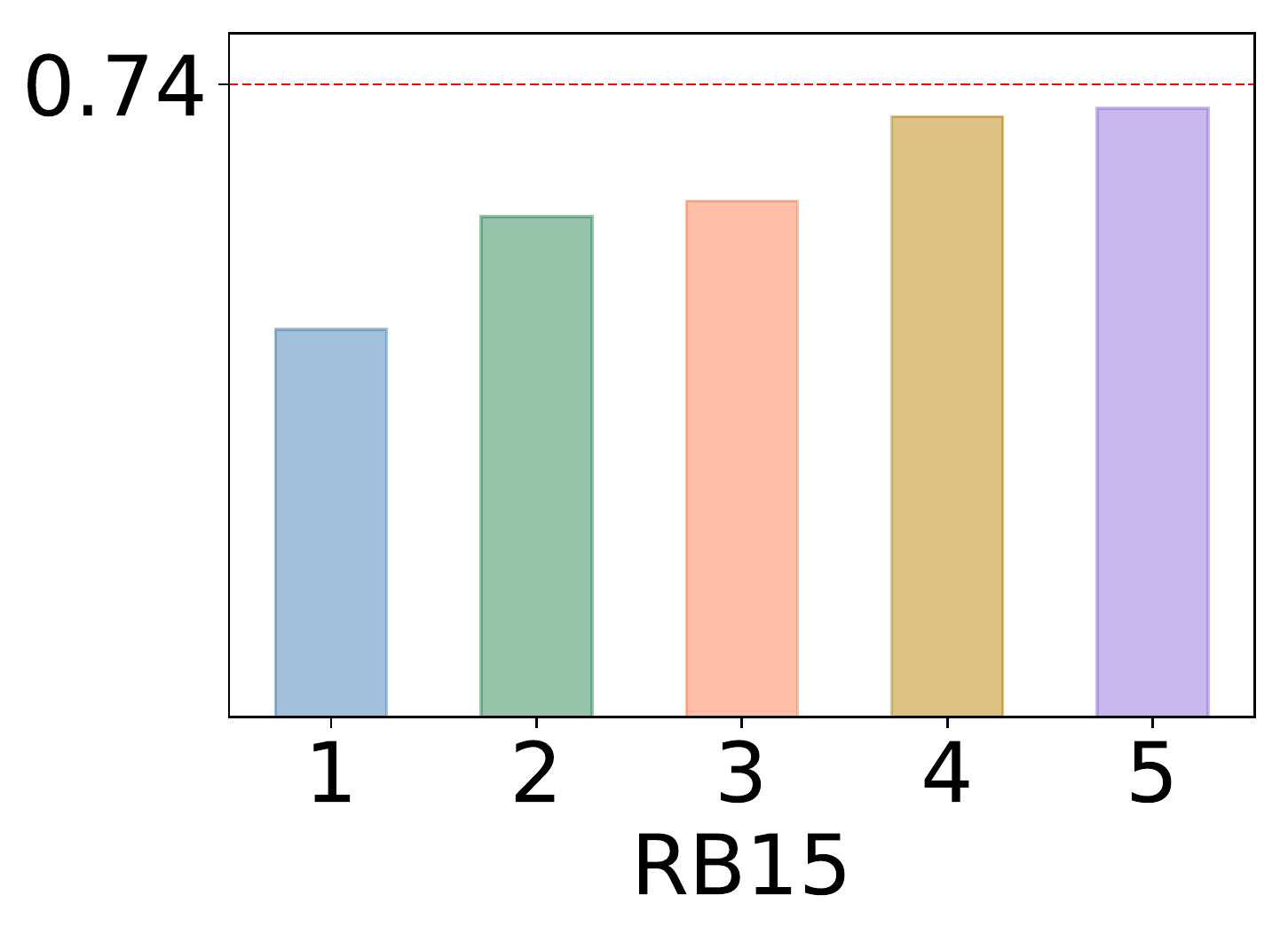}
  
\endminipage\hfill
\minipage{0.124\textwidth}%
  \includegraphics[width=\linewidth]{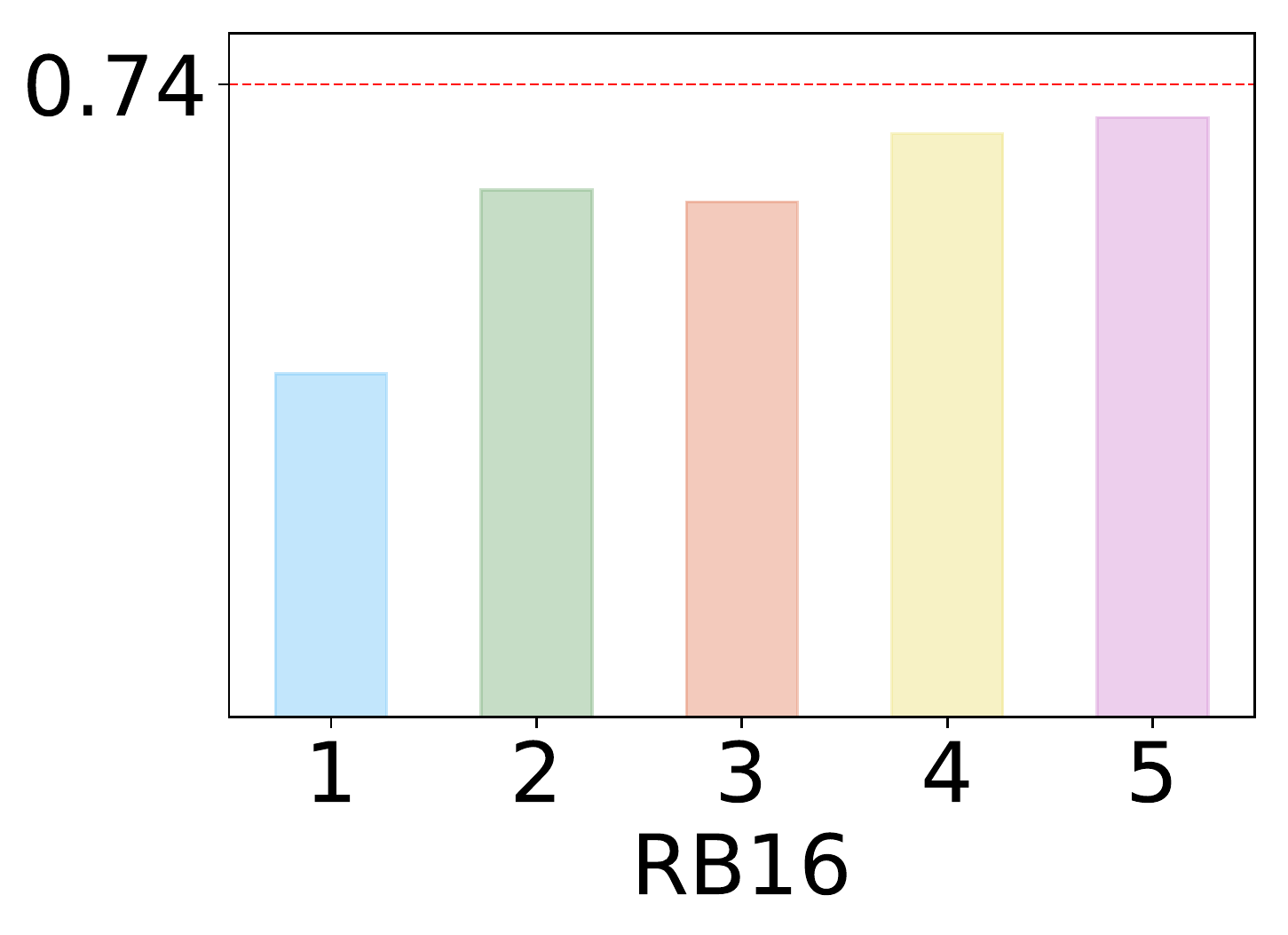}
  
\endminipage\hfill
\caption{Accuracy levels when choosing different $D_r$ values in the butterfly unit for all of its 16 possible locations in ResNet-50 (i.e. after RB1 to RB16). In this figure, only the results corresponding to $D_r$ values of 1-5 are presented. However, RB14, RB15, and RB16 require the minimum $D_r$ of 10 to maintain the accuracy of the proposed method at or above 74\% (less than 2\% accuracy loss).
}
\label{fig:accuracy_results}
\end{figure*}

\subsection{Latency and Energy Improvements}
The accuracy of ResNet-50 model for miniImageNet dataset without the butterfly unit is 76\%. We refer to this accuracy as the target accuracy. The accuracy results of the proposed method by placing the butterfly unit after each of the 16 residual blocks are demonstrated in Fig.~\ref{fig:accuracy_results}. According to Fig.~\ref{fig:accuracy_results}, as we increase the number of channels in the reduction unit, accuracy improves but larger feature tensors are needed to be transferred to the cloud. By assuming an acceptable error of 2\% compared to the target accuracy, placing the butterfly unit after residual blocks 1-3, 4-7, 8-13, and 14-16, requires the $D_r$ of 1, 2, 5, and 10, in order to maintain the accuracy of the proposed method at or above 74\% (less than 2\% accuracy loss), respectively.

\begin{table*}[ht]
	\caption{The End-to-End Latency, mobile energy consumption, and offloaded data size for different partition points in ResNet-50 using the proposed method 
	} 
	\centering 
	\begin{tabular}{|c|c|c|c|c|c|c|c|c|c|c|c|c|c|c|c|c|} \hline
	\textbf{Layer} & RB1 &  RB2 &  RB3 &  RB4 &  RB5 &  RB6 &  RB7 &  RB8 &  RB9 &  RB10 &  RB11 &  RB12 &  RB13 &  RB14 &  RB15 & RB16 \\
	\hline
	\textbf{Offloaded Data (KB)} & 3.1 &  3.1 &  3.1 &  1.6 &  1.6 &  1.6 &  1.6 &  1.0 &  1.0 &  1.0 & 1.0 & 1.0 & 1.0 &  0.5 &  0.5 &  0.5 \\
	\hline
	\textbf{Latency 3G (ms)} & 23.7 &	24.7 &	25.6 &	15.0 &	15.9 &	16.8 &	17.7 &	\optimalthreeglatency &	15.4 &	16.2 &	17.1 &	17.9 &	18.8 &	16.1 &	17.1 &	17.9 \\
	\hline
	\textbf{Energy 3G (mJ)} & 21.6 & 22.4 &	23.3 &	13.7 &	14.4 &	15.4 &	16.2 &	\optimalthreegenergy &	13.9 &	14.7 &	15.5 &	16.4 &	17.2 &	14.8 &	15.7 &	16.6\\
	\hline
	\textbf{Latency 4G (ms)} & \optimalfourglatency & 6.1 & 6.9 & 5.8 &	6.7 & 7.6 & 8.5 & 8.6 &	9.6 & 10.5 &	11.2 &	12.1 &	13.1 &	13.1 & 14.2 & 15.1\\
	\hline
	\textbf{Energy 4G (mJ)} & \optimalfourgenergy &	11.6 &	13.2 &	10.9 &	12.7 &	14.3 &	15.9 &	12.6 &	13.1 &	14.3 &	15.2 &	16.3 &	17.0 &	14.4 &	16.8 &	17.2\\
	\hline
	\textbf{Latency Wi-Fi (ms)} & \optimalwifilatency &	3.3	& 4.1 &	4.3 &	5.2 &	6.1 &	7.0 &	7.7 &	8.6 &	9.4 &	10.7 &	11.1 &	12.2 &	12.9 &	13.8 &	14.7\\
	\hline
	\textbf{Energy Wi-Fi (mJ)} & \optimalwifienergy & 6.8 &	8.7 &	9.1 &	11.2 &	13.1 &	14.9 &	12.1 &	12.7 &	13.9 &	14.8 &	15.5 &	16.3 &	14.1 &	16.1 &	16.6\\
	\hline
\end{tabular}
\label{table:layer_split} 
\end{table*}

\begin{table*}[ht]
\caption{Comparison of the proposed method with mobile-only and cloud-only approaches}
  \centering
\label{tab:lateny_energy_results}
\begin{tabular}{|c|c|c|c|c|c|c|c|}
\hline
\multicolumn{2}{|c|}{Setup}   & Latency (ms) & Energy (mJ) & Butterfly Unit Location & Offloaded Data (B) & Accuracy\\
\hline
\multirow{1}{*}{Mobile-only}  & -   &   15.7    & 20.5 & - & 0  & 76.1 \\
\hline
\multirow{3}{*}{Cloud-only}  & 3G   &   1101    & 1047.4 & - & 150528 & 76.1 \\
                         & 4G  &   208.4    & 528.3 & - & 150528 & 76.1 \\
                          & Wi-Fi  &   98.1   & 342.1  & - & 150528 & 76.1 \\
\hline
\multirow{3}{*}{Collaborative}  & 3G   &   14.3    & 13.1 & After RB8 & 980  & 74.0\\
                         & 4G  &  5.2    & 9.8 & After RB1 & 3136 & 74.1\\
                          & Wi-Fi  &   2.4    & 4.8  & After RB1 & 3136  & 74.1\\
\hline
\end{tabular}
\end{table*}
Table~\ref{table:layer_split} presents the latency and mobile energy consumption associated with placing the butterfly unit with proper $D_r$ size (with the accuracy loss less than 2\%) after each residual block, for different wireless networks when there is no congestion in the mobile, cloud, and wireless network. Table~\ref{tab:lateny_energy_results} shows the selected partition points by our algorithm for the goal of minimum end-to-end latency and mobile energy consumption, while the acceptable 2\% accuracy loss is reached, across three different wireless configurations (3G, 4G, and Wi-Fi) and when there is no congestion on the mobile, cloud, and wireless network (These selected partitions are also highlighted in Table~\ref{table:layer_split}). Note that the best partitioning for the goal of minimum end-to-end latency is the same as the best partitioning for the goal of minimum mobile energy consumption in each wireless network settings. This is mainly due to the fact that end-to-end latency and mobile energy consumption are proportional to each other since the dominant portion of both of them are associated with the wireless transfer overheads of the intermediate feature tensor. Obtained results for cloud-only and mobile-only approaches are also provided in Table~\ref{tab:lateny_energy_results}.

\textbf{Latency Improvement} - As demonstrated in Table~\ref{tab:lateny_energy_results}, using our proposed method, the end-to-end latency achieves 77$\times$, 40$\times$, 41$\times$ improvements over the cloud-only approach in 3G, 4G, and Wi-Fi networks, respectively.

\textbf{Energy Improvement} - As demonstrated in Table~\ref{tab:lateny_energy_results}, using our proposed method, the mobile energy consumption achieves 80$\times$, 54$\times$, and 71$\times$ improvements over the cloud-only approach in 3G, 4G, and Wi-Fi networks, respectively.

In the case of 4G and Wi-Fi, the mobile device is only required to compute DNN layers upon the completion of the RB1 and the reduction unit. In the case of 3G, the mobile device should compute DNN layers upon the completion of the RB8 and the reduction unit.


\subsection{Server Load Variations}
Data centers typically experience fluctuating load patterns. High server utilization leads to increased service times for DNN queries. Using our proposed method, by training multiple DNNs split on different layers and storing corresponding partitions in the mobile and cloud, the best model can be selected at run-time by the mobile, based on the current server load level, by periodically pinging the server during the mobile idle period. This leads to avoiding long latencies of DNN queries caused by high user demands. If the server is congested, we can move the partition point into deeper layers which this pushes more of the workload towards the mobile device. As a result, the computation load of the mobile device will increase. In summary, depending on the server load, the partition point can be changed while preserving the accuracy and still offloading less data than raw input.

Consequently, this new compute paradigm not only reduces the end-to-end latency and mobile energy consumption but also reduces the workload required on the data center, leading to the shorter query service time and higher query throughput. 

\subsection{Comparison to Other Feature Compression Techniques}
In the collaborative intelligence works which have considered the compression of intermediate features before uploading them to the cloud, the obtained compression ratios are significantly less compared to our work. For instance, as reported in \cite{choi2018deep}, the maximum achieved compression ratio is reported as 3.3$\times$. However, with the proposed trainable butterfly unit, we achieve up to 256$\times$ compression ratio when the butterfly unit is placed after RB1 (in which the channel size is reduced from 256 to 1). This shows that in collaborative intelligence framework, the compression using the proposed learnable butterfly unit can significantly perform better than traditional compressors.
\section{Conclusion and Future Work} \label{section.conclusion}
As the core component of today's intelligent services, DNNs have been traditionally executed in the cloud. 
Recent studies have shown that the latency and energy consumption of DNNs in mobile applications can be considerably reduced using collaborative intelligence framework, where the inference network is divided between the mobile and cloud and intermediate features computed on the mobile device are offloaded to the cloud instead of the raw input data of the network, reducing the size of the data needed to be sent to the cloud.
With these insights, in this work, we develop a new partitioning scheme that creates a bottleneck in a neural network using the proposed butterfly unit, which alleviates the communication costs of feature transfer between the mobile and cloud to a greater extent. It can adapt to any DNN architecture, hardware platform, wireless connections, and mobile and server load levels, and selects the best partition point for the minimum end-to-end latency and/or mobile energy consumption at run-time. The new network architecture, including the introduced butterfly unit after a selected layer of the underlying deep model, is trained end-to-end. Our proposed method, across different wireless networks, achieves on average 53$\times$ improvements for end-to-end latency and 68$\times$ improvements for mobile energy consumption compared to the status quo cloud-only approach for ResNet-50, while the accuracy loss is less than 2\%.

As a future work, the extent of reduction in the feature data size which is transferred between the mobile and cloud can be explored. Furthermore, the efficacy of the proposed method can be investigated for different DNN architectures and mobile/server load variations. Additionally, collaborative intelligence frameworks by considering the advent of revolutionary 5G technology can be studied.
\bibliographystyle{IEEEtran}
\bibliography{references.bib}

\begin{thebibliography}{10}
\providecommand{\url}[1]{#1}
\csname url@samestyle\endcsname
\providecommand{\newblock}{\relax}
\providecommand{\bibinfo}[2]{#2}
\providecommand{\BIBentrySTDinterwordspacing}{\spaceskip=0pt\relax}
\providecommand{\BIBentryALTinterwordstretchfactor}{4}
\providecommand{\BIBentryALTinterwordspacing}{\spaceskip=\fontdimen2\font plus
\BIBentryALTinterwordstretchfactor\fontdimen3\font minus
  \fontdimen4\font\relax}
\providecommand{\BIBforeignlanguage}[2]{{%
\expandafter\ifx\csname l@#1\endcsname\relax
\typeout{** WARNING: IEEEtran.bst: No hyphenation pattern has been}%
\typeout{** loaded for the language `#1'. Using the pattern for}%
\typeout{** the default language instead.}%
\else
\language=\csname l@#1\endcsname
\fi
#2}}
\providecommand{\BIBdecl}{\relax}
\BIBdecl

\bibitem{krizhevsky2012imagenet}
A.~Krizhevsky, I.~Sutskever, and G.~E. Hinton, ``Imagenet classification with
  deep convolutional neural networks,'' in \emph{Advances in neural information
  processing systems}, 2012, pp. 1097--1105.

\bibitem{Resnet}
\BIBentryALTinterwordspacing
K.~He, X.~Zhang, S.~Ren, and J.~Sun, ``Deep residual learning for image
  recognition,'' \emph{CoRR}, vol. abs/1512.03385, 2015. [Online]. Available:
  \url{http://arxiv.org/abs/1512.03385}
\BIBentrySTDinterwordspacing

\bibitem{girshick2016region}
R.~Girshick, J.~Donahue, T.~Darrell, and J.~Malik, ``Region-based convolutional
  networks for accurate object detection and segmentation,'' \emph{IEEE
  transactions on pattern analysis and machine intelligence}, vol.~38, no.~1,
  pp. 142--158, 2016.

\bibitem{hinton2012deep}
G.~Hinton, L.~Deng, D.~Yu, G.~E. Dahl, A.-r. Mohamed, N.~Jaitly, A.~Senior,
  V.~Vanhoucke, P.~Nguyen, T.~N. Sainath \emph{et~al.}, ``Deep neural networks
  for acoustic modeling in speech recognition: The shared views of four
  research groups,'' \emph{IEEE Signal processing magazine}, vol.~29, no.~6,
  pp. 82--97, 2012.

\bibitem{mikolov2013distributed}
T.~Mikolov, I.~Sutskever, K.~Chen, G.~S. Corrado, and J.~Dean, ``Distributed
  representations of words and phrases and their compositionality,'' in
  \emph{Advances in neural information processing systems}, 2013, pp.
  3111--3119.

\bibitem{choi2018deep}
H.~Choi and I.~V. Bajic, ``Deep feature compression for collaborative object
  detection,'' \emph{arXiv preprint arXiv:1802.03931}, 2018.

\bibitem{eshratifar2018jointdnn}
A.~E. Eshratifar, M.~S. Abrishami, and M.~Pedram, ``Jointdnn: an efficient
  training and inference engine for intelligent mobile cloud computing
  services,'' \emph{arXiv preprint arXiv:1801.08618}, 2018.

\bibitem{EshratifarGLS}
\BIBentryALTinterwordspacing
A.~E. Eshratifar and M.~Pedram, ``Energy and performance efficient computation
  offloading for deep neural networks in a mobile cloud computing
  environment,'' pp. 111--116, 2018. [Online]. Available:
  \url{http://doi.acm.org/10.1145/3194554.3194565}
\BIBentrySTDinterwordspacing

\bibitem{kang2017neurosurgeon}
Y.~Kang, J.~Hauswald, C.~Gao, A.~Rovinski, T.~Mudge, J.~Mars, and L.~Tang,
  ``Neurosurgeon: Collaborative intelligence between the cloud and mobile
  edge,'' \emph{ACM SIGPLAN Notices}, vol.~52, no.~4, pp. 615--629, 2017.

\bibitem{grulich2018collaborative}
P.~M. Grulich and F.~Nawab, ``Collaborative edge and cloud neural networks for
  real-time video processing,'' \emph{Proceedings of the VLDB Endowment},
  vol.~11, no.~12, pp. 2046--2049, 2018.

\bibitem{chen2018intermediate}
Z.~Chen, W.~Lin, S.~Wang, L.~Duan, and A.~C. Kot, ``Intermediate deep feature
  compression: the next battlefield of intelligent sensing,'' \emph{arXiv
  preprint arXiv:1809.06196}, 2018.

\bibitem{choi2018near}
H.~Choi and I.~V. Bajic, ``Near-lossless deep feature compression for
  collaborative intelligence,'' \emph{arXiv preprint arXiv:1804.09963}, 2018.

\bibitem{JetsonTX2}
``{Jetson TX2 Module},''
  \url{https://developer.nvidia.com/embedded/buy/jetson-tx2}, 2018.

\bibitem{INA226}
``{INA Current/Power Monitor},'' \url{http://www.ti.com/product/INA226}.

\bibitem{MobNet}
``{State of Mobile Networks in USA},''
  \url{https://opensignal.com/reports/2017/08/usa/state-of-the-mobile-network},
  2017.

\bibitem{Speedtest}
``{United States Speedtest Market Report},''
  \url{http://www.speedtest.net/reports/united-states/}, 2017.

\bibitem{4GLTE}
J.~Huang, F.~Qian, A.~Gerber, Z.~M. Mao, S.~Sen, and O.~Spatscheck, ``A close
  examination of performance and power characteristics of 4g lte networks,'' in
  \emph{Proceedings of the 10th International Conference on Mobile Systems,
  Applications, and Services}, ser. MobiSys '12.\hskip 1em plus 0.5em minus
  0.4em\relax New York, NY, USA: ACM, 2012, pp. 225--238.

\bibitem{TensorRT}
``{NVIDIA TensorRT},''
  \url{https://docs.nvidia.com/deeplearning/sdk/tensorrt-api/index.html}, 2018.

\bibitem{cuDNN}
\BIBentryALTinterwordspacing
S.~Chetlur, C.~Woolley, P.~Vandermersch, J.~Cohen, J.~Tran, B.~Catanzaro, and
  E.~Shelhamer, ``cudnn: Efficient primitives for deep learning,'' \emph{CoRR},
  vol. abs/1410.0759, 2014. [Online]. Available:
  \url{http://arxiv.org/abs/1410.0759}
\BIBentrySTDinterwordspacing

\bibitem{deepcompression}
\BIBentryALTinterwordspacing
S.~Han, H.~Mao, and W.~J. Dally, ``Deep compression: Compressing deep neural
  network with pruning, trained quantization and huffman coding,'' \emph{CoRR},
  vol. abs/1510.00149, 2015. [Online]. Available:
  \url{http://arxiv.org/abs/1510.00149}
\BIBentrySTDinterwordspacing

\bibitem{ImageNet}
J.~Deng, W.~Dong, R.~Socher, L.~Li, K.~Li, and L.~Fei-Fei, ``Imagenet: A
  large-scale hierarchical image database,'' in \emph{2009 IEEE Conference on
  Computer Vision and Pattern Recognition}, June 2009, pp. 248--255.

\bibitem{slee2007thrift}
\BIBentryALTinterwordspacing
M.~Slee, A.~Agarwal, and M.~Kwiatkowski, ``{Thrift: Scalable cross-language
  services implementation},'' \emph{Facebook White Paper}, vol.~5, 2007.
  [Online]. Available:
  \url{http://thrift.apache.org/static/files/thrift-20070401.pdf}
\BIBentrySTDinterwordspacing

\bibitem{jousha_difficulty}
\BIBentryALTinterwordspacing
X.~Glorot and Y.~Bengio, ``Understanding the difficulty of training deep
  feedforward neural networks,'' in \emph{Proceedings of the Thirteenth
  International Conference on Artificial Intelligence and Statistics}, ser.
  Proceedings of Machine Learning Research, Y.~W. Teh and M.~Titterington,
  Eds., vol.~9.\hskip 1em plus 0.5em minus 0.4em\relax Chia Laguna Resort,
  Sardinia, Italy: PMLR, 13--15 May 2010, pp. 249--256. [Online]. Available:
  \url{http://proceedings.mlr.press/v9/glorot10a.html}
\BIBentrySTDinterwordspacing

\bibitem{miniImageNet}
\BIBentryALTinterwordspacing
O.~Vinyals, C.~Blundell, T.~P. Lillicrap, K.~Kavukcuoglu, and D.~Wierstra,
  ``Matching networks for one shot learning,'' \emph{CoRR}, vol.
  abs/1606.04080, 2016. [Online]. Available:
  \url{http://arxiv.org/abs/1606.04080}
\BIBentrySTDinterwordspacing

\end{thebibliography}

\end{document}